\documentclass[12pt,a4paper]{article}
\pdfoutput=1
\usepackage{amsmath,amssymb,amsfonts,geometry,enumerate,graphicx,epsfig,cite}
\usepackage{fullpage}
\usepackage{centernot,stmaryrd}
\newcommand{\beq}{\begin{eqnarray}}
\newcommand{\eeq}{\end{eqnarray}}
\newcommand{\bea}{\begin{eqnarray}}
\newcommand{\eea}{\end{eqnarray}}
\newcommand{\bec}{\begin{center}}
\newcommand{\eec}{\end{center}}
\numberwithin{equation}{section}

\newcommand{\be}{\begin{equation}}
\newcommand{\ee}{\end{equation}}

\renewcommand{\hat}{\widehat}
\renewcommand{\tilde}{\widetilde}
\renewcommand{\epsilon}{\varepsilon}

\def\ben{\begin{equation}}
\def\een{\end{equation}}
\def\beq{\begin{equation}}
\def\eeq{\end{equation}}
\def\bea{\begin{eqnarray}}
\def\eea{\end{eqnarray}}

\newcommand{\rhot}{\widetilde\rho}
\newcommand{\bt}{\widetilde b}
\def\ss{Sakai-Sugimoto\ }
\def\h{H^{-\frac{1}{2}}}
\def\k{H^{\frac{3}{2}}}

\begin{document}

\title{
\begin{flushright}\ \vskip -2cm {\small{\em DCPT-13/31}}\end{flushright}
\vskip 2cm The Sakai-Sugimoto soliton}
\author{
Stefano Bolognesi and Paul Sutcliffe\\[10pt]
{\em \normalsize Department of Mathematical Sciences,
Durham University, Durham DH1 3LE, U.K.}\\[10pt]
{\normalsize Email: \quad  
s.bolognesi@durham.ac.uk \quad\&\quad\  p.m.sutcliffe@durham.ac.uk}
}
\date{September 2013}
\maketitle
\begin{abstract}
The Sakai-Sugimoto model is the preeminent example of a string theory
description of holographic QCD, in which baryons correspond to topological
solitons in the bulk. Here we investigate the validity of various
approximations of the Sakai-Sugimoto soliton 
that are used widely to study the properties of holographic baryons.
These approximations include the flat space self-dual instanton,
a linear expansion in terms of eigenfunctions in the holographic
direction and an asymptotic power series at large radius.
These different approaches have produced contradictory results
in the literature regarding properties of the baryon, such as
relations for the electromagnetic form factors.
Here we determine the regions of validity of these various approximations 
and show how to relate different approximations in 
contiguous regions of applicability.
This analysis clarifies the source of the contradictory results in the 
literature and resolves some outstanding issues,
 including the use of the flat space self-dual instanton,
the detailed properties of the asymptotic soliton tail, and the 
role of the UV cutoff introduced in previous investigations. 
A consequence of our analysis is the discovery of a new large scale,
that grows logarithmically with the 't Hooft coupling, at
which the soliton fields enter a nonlinear regime.
Finally, we provide the first numerical computation of the Sakai-Sugimoto 
soliton and demonstrate that the numerical results support our analysis.
\end{abstract}

\newpage

\section{Introduction}\quad
There are still some unresolved puzzles regarding aspects of bulk solitons in 
holographic models of QCD. These include the validity of the flat space
self-dual instanton used in the \ss model and the large distance behaviour of 
the electromagnetic form factors of the baryon. 
In this paper we shall address these issues and provide analytic resolutions
that are confirmed by numerical investigations. 

The cornerstone of all models of baryons in holographic QCD is that 
solitons in the bulk correspond to Skyrmions on the boundary. 
This correspondence was first observed by Atiyah and Manton 
\cite{Atiyah:1989dq} in four-dimensional Euclidean space, 
where the bulk soliton is the self-dual Yang-Mills instanton. The
correspondence can be formulated as a flat space version of holography
\cite{Sutcliffe:2010et}.
Holographic QCD differs from the Atiyah-Manton approach in that 
 spacetime is curved with AdS-like behaviour and a five-dimensional
Chern-Simon term is included that generates an abelian electric charge 
for the soliton. Here AdS-like means that the curvature is negative and 
there is a conformal boundary. 
The combination of the curvature of spacetime and the electromagnetic 
repulsion provides a stability that fixes the size of the soliton.
These features are common to all models of holographic QCD, 
whether bottom-up or top-down.

Top-down approaches are derived from a string embedding and the 
\ss  model \cite{Sakai:2004cn,Sakai:2005yt}
 is the prototypical example for top-down AdS/QCD models. 
In these models, the validity of the supergravity approximation 
requires working with a large number of colours $N_c$ and a large value 
of the 't Hooft coupling $\lambda$. 
Although $N_c$ is just an overall multiplicative factor in the action, 
and thus irrelevant at the classical level, $\lambda$ plays
a vital role for the classical soliton, as it controls the ratio 
between the Yang-Mills and Chern-Simons terms. 
In particular, for large $\lambda$ the size of the soliton 
becomes parametrically small with respect to the curvature scale. 
This suggests that most of the energy density of the soliton is 
concentrated in a small region of space, where the effect of the 
curvature has little influence on the fields of the soliton. 
This motivates the approach used in \cite{Hong:2007kx,Hata:2007mb}, where the
soliton is approximated by the flat space self-dual Yang-Mills instanton, 
with a size determined by minimization of the energy function
on the instanton moduli space that results by restricting the full energy 
functional to the space of self-dual instanton fields. 
Note that this approximation is based on the assumption that 
the curvature and Chern-Simons term do not significantly alter the 
soliton fields, even though they are crucial in determining its size. 
We shall put this assumption to the test by numerically computing
the \ss soliton and comparing it to the self-dual instanton.
Furthermore, we show how to improve the self-dual instanton approximation 
via a simple generalization that maintains the $SO(4)$ symmetry of 
the instanton but introduces a more general profile function.  

The soliton properties at large distance, and consequently the baryon 
electromagnetic form factors of the dual theory, 
have been calculated by expanding the self-dual instanton tail at the 
linear level and then extending this linear solution into the curved space
 at large distance from the core 
\cite{Hashimoto:2008zw,Hong:2007dq,Kim:2008pw}. 
This approach relies on the fact that, for a small soliton, there is a 
region from the soliton core to the curvature scale 
in which the soliton is essentially in a linear regime and the curvature 
effects remain negligible. 
The result of this linear analysis is that the baryon density, 
and consequently all the electromagnetic form factors (including those
of exited baryons obtained from a zero mode quantization) 
are exponentially suppressed at large distance.
This is in contrast to the situation for other models, 
including the standard Skyrme model with massless pions, 
where the baryon density has an algebraic decay.

Bottom-up approaches are equally good toy-models for AdS/QCD, 
as long as they incorporate the features of confinement and chiral 
symmetry breaking. 
These models are relieved of the requirement of a string theory embedding, 
so there is a free choice of any AdS-like metric (provided it has a 
conformal boundary in the UV) and $\lambda$ need not be small.
The Pomarol-Wulzer model \cite{Pomarol:2008aa} is an example in this
category, where the metric is a slice of AdS$_5$ with a finite IR boundary 
at which left and right gauge fields are subject to
matching conditions that mimic the salient features of the \ss model. 
Numerical computations of the Pomarol-Wulzer soliton 
have been performed at a value of the coupling that 
is of order one (where the soliton size is comparable to the 
curvature scale of the AdS$_5$ slice), together with
an asymptotic power series at large radius 
\cite{Panico:2008it}.
These results show that the baryon form factors have an algebraic decay, 
as in the Skyrme model, and not an exponential decay.

Cherman, Cohen and Nielsen \cite{Cherman:2009gb} have described model 
independent relations for the baryon form factors at large distance.  
These relations are satisfied by the baryon form factors computed in the 
Skyrme model and the Pomarol-Wulzer model but not by those
of the \ss model obtained from the linear analysis. 
The exponential decay of the soliton fields in the \ss model 
lies at the heart of this failure.
Later, Cherman and Ishii \cite{Cherman:2011ve}
adapted the large radius expansion in \cite{Panico:2008it} to the
\ss model and found that the form factors have an algebraic decay
and indeed satisfy the model independent relations, contradicting the 
earlier result of the linear analysis.
However, their approach required the introduction of a UV cutoff and 
problems arise in attempting to remove this cutoff, so it is not clear 
which of the conflicting results is correct. 
Very recently, a preprint has appeared in which the large radius 
expansion has been applied to a general metric \cite{Colangelo:2013pxk}
and a conclusion drawn regarding the UV cutoff introduced
into the \ss model. 
We shall comment on this conclusion in section \ref{sec-CI}, where
we derive the correct procedure for removing the UV cutoff. 

The contradictory conclusions described above raise a number of issues 
and questions concerning the use and validity of the various approximations 
and approaches. In fact, several candidates have been suggested for the 
source of the disagreement.
One possibility is that the use of the flat space self-dual instanton
in the \ss model is at the root of the problem.
The validity of this approximation has never been tested, 
either numerically or analytically, and one may worry about a mechanism 
that allows the curvature and Chern-Simons term to stabilize the instanton 
size without altering the form of its fields. 
We shall test the use of the instanton approximation, firstly by
introducing a generalization that allows some deformation of the
instanton fields, and secondly via direct numerical computation of the 
\ss soliton. Our results strongly support
 the validity of the self-dual instanton approximation for large 't Hooft 
coupling.

Another possibility is that either the linear expansion in
\cite{Hashimoto:2008zw} 
or the large radius expansion in \cite{Cherman:2011ve}
are not valid in the \ss model.
In fact, we shall show that both approaches are valid but they are applicable
in different regions of space.
The contradictory results concerning the soliton tail, and consequently 
the baryon form factors, is a result of applying the linear expansion
in an inappropriate region.
The resolution of all the discrepancies in the literature
resides in the existence of a new scale. 
This is a large scale that grows logarithmically with the 't Hooft
coupling and is therefore much larger than both the radius of curvature
and the size of the small instanton. 
The linear expansion should be thought of as an expansion
in $\lambda^{-1}$, where the first term solves the linearised 
field equations. However, higher order terms are larger than the first
order term both at the small instanton scale, which is of order
$\lambda^{-{1}/{2}},$  and crucially at the new large scale of order
$\log{\lambda}$. The fundamental property of the system is that there
is a transition from a linear to a nonlinear regime at large distance.
The existence of this new large scale explains the discrepancy over the 
form factor computations, which depend on the fall-off of the soliton tail.
The vital observation is that the large $\lambda$ and large
radius limits do not commute. The crucial terms with algebraic decay
are suppressed by additional powers of $\lambda^{-1}$ in comparison to
the terms with exponential decay, so the algebraic decay 
is only evident at the large scale of order $\log {\lambda}$.
 
The outline of this paper is as follows.
In section \ref{sec-ss} we review the main aspects of the \ss model.
In section \ref{radialandselfdual} we discuss the flat space self-dual
instanton approximation and our radial generalization.
Section \ref{tail} is devoted to the calculation of the tail properties
of the soliton, and in particular a determination of the regions of
validity of alternative approximations. By comparing these different
approximations we are able to relate them to each other and hence
predict the emergence of the new large scale.
A numerical computation of the \ss soliton is described in section
\ref{numerics}, where the numerical results are shown to support our
analytic findings. Finally, some concluding remarks are made in section 
\ref{conclusion}.

\section{The Sakai-Sugimoto model}\quad\label{sec-ss}
Consider a five-dimensional spacetime with a warped metric of the
form
\beq
\label{metric}
ds^2 = H(z) \,dx_{\mu} dx^{\mu} +  \frac{1}{H(z)}dz^2.
\eeq 
Here $x_\mu,$ with $\mu=0,1,2,3,$ are the coordinates of four-dimensional
Minkowski spacetime and $z$ is the spatial coordinate in the additional
holographic direction. 
The signature is $(-,+,+,+,+)$. 

A class of spacetimes that are particularly relevant for 
holographic baryons corresponds to the choice
\beq
\label{class}
H = \left(1+\frac{z^2}{L^2}\right)^p,
\eeq
where $L$ and $p$ are positive constants, with 
the former setting a curvature length scale.
In this paper we focus on the Sakai-Sugimoto model 
\cite{Sakai:2004cn,Sakai:2005yt}, which
corresponds to the choice $p=\frac{2}{3}$.
For general $p$ 
the scalar curvature of the metric, after setting the length scale $L=1$, is
\bea
R = - 4 H^{-3/4} \left(H^{3/4} H' \right)' =
 -\frac{4p \left( 2 + (7p -2 ) z^2  \right)}{\left(1+z^2\right)^{2-p} }.
\eea
This formula shows that the value of $p$ is crucial in 
determining the qualitative features of the spacetime.
For $p\le 1$ the curvature is finite as $z\to\infty.$
For $p=1$ the spacetime is asymptotically AdS$_5$ with constant negative 
curvature $-20$.
For $p > \frac{2}{7}$ the curvature is negative for all $z$ and   
for $p> \frac{1}{2}$ the theory has a conformal boundary. 
In the case of a conformal boundary it is often useful to introduce 
conformal coordinates
\beq
ds^2 = H(z(u)) \left(dx_{\mu} dx^{\mu} + du^2\right),
\eeq
where $u$ solves the equation $du/dz =  1/H(z).$ 
For large $z$ the asymptotic behaviour is 
$u(z) \simeq c_1 + c_2/z^{2p-1},$ for some constants $c_1$ and $c_2.$
Thus $u \to c_1$ as $z \to \infty$, revealing the conformal boundary.

Given the above properties, we refer to the metric as 
AdS-like if $p\in(\frac{1}{2},1],$ since there is then a conformal boundary and
the curvature is negative and finite.
The Sakai-Sugimoto model is a generic example with $p=\frac{2}{3}.$
Unless otherwise specified, from now on we will fix the values $L=1$
and $p=\frac{2}{3},$ though occasionally we will reintroduce these 
constants to indicate the more general dependence.

The \ss model is a $U(2)$ gauge theory in the five-dimensional 
spacetime introduced above. Our index notation is that
uppercase indices include the holographic direction whilst lowercase 
indices exclude this additional dimension. Furthermore, 
greek indices include the time coordinate whilst latin indices 
(excluding $z$) run over the spatial coordinates. Thus, for example,
\be
\Gamma,\Delta,\ldots=0,1,2,3,z, \quad
\mu,\nu,\ldots=0,1,2,3, \quad
I,J,\ldots=1,2,3,z, \quad
i,j,\ldots=1,2,3.
\ee
To fix conventions, the gauge potential ${\cal A}_\Gamma$ 
is hermitian and under a gauge transformation, $G\in U(2),$ it transforms as
$
{\cal A}_\Gamma\mapsto G {\cal A}_\Gamma G^{-1}+i(\partial_\Gamma G) G^{-1}.
$
The associated field strength is  
${\cal F}_{\Gamma\Delta}=\partial_\Gamma {\cal A}_\Delta
-\partial_\Delta {\cal A}_\Gamma+i[{\cal A}_\Gamma,{\cal A}_\Delta]$
and the covariant derivative is 
$D_{\Gamma}\heartsuit = \partial_{\Gamma}\heartsuit + i  [A_{\Gamma},\heartsuit].$
The action is the sum of a Yang-Mills term and a $U(2)$ Chern-Simons term
\bea
\label{actionymcs}
{\cal S} = -\frac{N_c \lambda}{216 \pi^3}   \int \sqrt{-g} \  \frac{1}{2} {\rm tr} \left({\cal F}_{\Gamma\Delta} {\cal F}^{\Gamma\Delta}\right)
\,d^4x\, dz 
 +  \frac{N_c}{24 \pi^2} \int 
\ \omega_5({\cal A})\,d^4x\, dz, 
\eea
where $g$ is the earlier warped metric with $p=\frac{2}{3}.$
The factors $N_c$ and $\lambda$ are respectively the number of colours and 
the 't Hooft coupling of the dual theory. 
Note that the number of colours acts just as a multiplicative factor and 
therefore plays a trivial role in the classical physics in the bulk. 
In particular, 
by keeping $\lambda$ fixed and taking the limit $N_c \to \infty$ we can 
always make any quantum corrections negligible.

Decomposing the $U(2)$ gauge potential into a sum of 
non-abelian $SU(2)$ and abelian $U(1)$ components 
\beq
{\cal A}_{\Gamma} = A_{\Gamma} + \frac{1}{2} \hat{A}_{\Gamma}, \qquad\qquad 
{\cal F}_{\Gamma} = F_{\Gamma} + \frac{1}{2} \hat{F}_{\Gamma},
\eeq
the $U(2)$ Chern-Simons term, up to a total derivative, is
\beq
 \frac{N_c}{24 \pi^2} \int 
\left( \frac{3}{8} \hat{A}_{\Gamma} {\rm tr}\,( {F}_{\Delta\Sigma}{ F}_{\Xi\Upsilon})   + \frac{1}{16} \hat{A}_{\Gamma} \, \hat{F}_{\Delta\Sigma} \hat{ F}_{\Xi\Upsilon} \right) \epsilon^{\Gamma\Delta\Sigma\Xi\Upsilon}
\,d^4x\,dz.  
\eeq
The action, conveniently rescaled,  becomes
\bea
S &=& \frac{ 216 \pi^3 }{N_c\lambda} {\cal S} \nonumber \\ &=& 
\int 
\left\{ - \frac{1}{4 H^{1/2}} {\hat{F}}_{\mu \nu }{\hat{F}}^{\mu \nu } - \frac{H^{3/2}}{2} {\hat{F}}_{\mu  z}{\hat{F}}^{\mu z } \right. 
  \left.    - \frac{1}{2 H^{1/2}} {\rm tr} \left({ F}_{\mu \nu}{ F}^{\mu \nu}\right ) - H^{3/2} {\rm tr} \left({F}_{\mu  z} {F}^{\mu  z}\right ) \right\}\,d^4x\,dz     \nonumber \\
&&  +   
\frac{1}{\Lambda}\int    \left(\hat{A}_{\Gamma} {\rm tr}\,( {F}_{\Delta\Sigma}{ F}_{\Xi\Upsilon}) + \frac{1}{6} \hat{A}_{\Gamma} \, \hat{F}_{\Delta\Sigma} \hat{ F}_{\Xi\Upsilon} \right) \, \epsilon^{\Gamma\Delta\Sigma\Xi\Upsilon}
\,d^4x\,dz,     
\eea
where the indices are now raised using the flat 5-dimensional 
Minkowski metric tensor $\eta_{\Gamma\Delta}$.
For convenience, in the above we have introduced the 
rescaled 't Hooft coupling 
\beq
\Lambda = \frac{8 \lambda}{27 \pi}. 
\eeq

As we are concerned with the static soliton solution of the theory,
from now on we shall restrict to the case of time independent fields.
The appropriate static ansatz is
\beq
{A}_0=0, \qquad A_I={A}_I(x_{J}), \qquad  
\hat{A}_0=\hat{A}_0(x_{J}), \qquad \hat{A}_{I} = 0,
\eeq
so that the abelian potential generates an electric field 
$\hat{F}_{I0}= \partial_I \hat{A}_0$.
The action restricted to static fields is then
\bea
\label{staticaction}
S &=& \int 
\left\{   \frac{1}{2 H^{1/2}} (\partial_i \hat{A}_0)^2 + 
\frac{H^{3/2}}{2} (\partial_z \hat{A}_0)^2 \right.
  \left. - \frac{1}{2 H^{1/2}} {\rm tr}  \left( { F}_{ij}^2 \right) 
- H^{3/2} {\rm tr}  \left( {F}_{i  z}^2 \right) \right\} 
\,d^4x\,dz  
    \nonumber \\
&&  +   \frac{1}{\Lambda}\int 
\hat{A}_0 \ {\rm tr}\,( {F}_{IJ}{ F}_{KL}) \,  \epsilon_{IJKL} 
\,d^4x\,dz. 
\eea

The static field equations that follow from the variation of this
action are
\bea
&& \frac{1}{H^{1/2}} D_jF_{ji}+D_z(H^{3/2}F_{zi})
=\frac{1}{\Lambda}\epsilon_{iJKL}F_{KL}
\partial_J\hat A_0
\label{feq1}
\\
&& H^{3/2}D_jF_{jz}=\frac{1}{\Lambda}\epsilon_{ijk}F_{jk}\partial_i\hat A_0
\label{feq2}
\\
&&\frac{1}{H^{1/2}}\partial_i\partial_i\hat A_0
+\partial_z(H^{3/2}\partial_z\hat A_0)=
\frac{1}{\Lambda} {\rm tr}\,(F_{IJ}F_{KL})\,\varepsilon_{IJKL}.
\label{feq3}
\eea

Baryon number is identified with the $SU(2)$ instanton number
of the soliton
\be
B=-\frac{1}{32\pi^2}\int {\rm tr}(F_{IJ}F_{KL})\,\varepsilon_{IJKL}
\,d^3x\, dz,
\ee
and the Chern-Simons coupling implies that the instanton charge 
density sources the abelian electric field.

For later computational purposes, it will be convenient to
rewrite the action 
by rearranging the terms as
\bea
\label{staticactionisolate}
S &=& 
\int 
\left\{  
\frac{ H^{3/2}}{2}\bigg( (\partial_{I} \hat{A}_0)^2
- {\rm tr}\left({F}_{IJ}^2\right)\bigg) 
 + \frac{1 - H^2}{2H^{1/2}}\bigg( (\partial_i \hat{A}_0)^2 
-{\rm tr} \left({F}_{ij}^2\right)\bigg)  \right\}
\,d^4x\, dz,
  \nonumber \\
&&  +   
\frac{1}{\Lambda}\int \hat{A}_0 \ {\rm tr}\,( {F}_{IJ}{ F}_{KL}) \, 
 \epsilon_{IJKL}\,d^4x\, dz. 
\eea
\newpage
\section{Radial and self-dual approximations}\quad
\label{radialandselfdual}
As we shall see, the static soliton solution of 
the field equations that follow from 
(\ref{staticaction}) is quite complicated.
Even for the single static soliton, symmetry reduction can only
reduce the field equations to coupled partial differential equations for five
functions of two variables, which then need to be solved numerically.
This approach will be described in detail in section \ref{numerics}, where we present
the results of the first numerical computation of the \ss soliton. 

The lack of an exact solution has motivated various approximate descriptions
of the soliton, some of which we shall discuss later. 
First we consider a new approximation, in which the fields are assumed to have
$SO(4)$ spherical symmetry. Because of the warp factor in the metric, such
an assumption is clearly incompatible with the true solution of the field
equations, so no exact solutions can be obtained in this way. 
However, we can certainly restrict the functional space to such
a set of symmetric trial fields and determine the fields that are
stationary points of the restricted action. The advantage of this approach
is that it reduces the problem to a single ordinary differential
equation, which is much easier to deal with than the full 
coupled partial differential equations. 
Furthermore, the radial approximation is a generalization of the self-dual
flat space instanton approximation that has been used heavily in 
previous studies, so we are able to further investigate this approximation
by examining how the radial approximation
compares to the self-dual approximation in the large $\Lambda$ limit.
The obvious disadvantage of the radial approximation 
is that it is unclear whether the approximate fields
provide a reasonable description of the true solution. Fortunately, our
later numerical solution will allows us to investigate this aspect too.
 
To specify the fields within the radial approximation we define
the coordinates $\rho\ge 0$ and $\theta\in [0,\pi]$ by 
\beq
\rho = \sqrt{x_1^2 + x_2^2 + x_3^2 + z^2}, \qquad 
z = \rho \cos{\theta}.
\eeq
The radial approximation involves two real profile functions
$a(\rho)$ and $b(\rho)$ and is given by
\beq
\label{radialansatz}
\hat{A}_0= a(\rho), \qquad\qquad
A_{I} = - {\sigma}_{IJ} x_{J} \, b(\rho),
\eeq
where $\sigma_{IJ}$ is the anti-symmetric 
't Hooft tensor defined in terms of the
Paul matrices $\sigma_i$ by
\beq
\sigma_{ij} = \epsilon_{ijk}\sigma_k, 
\qquad \sigma_{zi} =  \sigma_i.
\eeq
The non-abelian field has the same $SO(4)$ symmetry as the 
self-dual instanton, but has a more general radial profile function.

The instanton charge density is
\bea
-\frac{1}{32\pi^2} {\rm tr}(F_{IJ}F_{KL})\,\varepsilon_{IJKL}
&=& \frac{3}{\pi^2} b(1-\rho^2 b)(2b + \rho b') \nonumber \\
&=&
\frac{1}{\pi^2\rho^3}\left(\frac{3}{2}(\rho^2 b)^2-(\rho^2 b)^3\right)'
\eea
yielding the instanton number 
\be
B=c^2(3-2c), \quad \mbox{where}\quad c=\lim_{\rho\to\infty}(\rho^2 b).
\ee
The requirement that $B=1$ therefore determines that $c=1,$ giving the
large $\rho$ behaviour 
\be
b=\frac{1}{\rho^2}
+{\cal O}\bigg(\frac{1}{\rho^4}\bigg).
\ee
In evaluating the action density of the radial field, 
the first term to consider is 
\beq
  {\rm tr} \left({ F}_{IJ}^2 \right)= 
12(2b+\rho b')^2+48b^2(1-b\rho^2)^2.
\eeq
The remaining term that is required is
\beq
  {\rm tr} \left( { F}_{i j}^2 \right)=
\frac{8}{\rho}\bigg(6\rho b^2+2b^3\rho(\rho^2+2z^2)(b\rho^2-2)+b'(b'\rho+4b)(\rho^2-z^2)\bigg).  
\eeq
Substituting these expressions into the action
(\ref{staticactionisolate}), writing $z=\rho\cos\theta$ and performing the 
angular integration over $\theta$ gives 
\bea
\label{funcs}
\frac{S}{2\pi^2} &=& \int \Big\{(P_1 - P_2+ P_3) a'^2   
-12 P_1\big((2b+\rho b')^2+4b^2(1-b\rho^2)^2\big) \nonumber\\
  &&
+8\big(6b^2P_2+2b^3\rho^2(P_2+2P_3)(b\rho^2-2)+b'(b'\rho+4b)\rho(P_2-P_3)\big)  
\Big\}\rho^3\,d\rho\,dt
 \nonumber \\
&&
-\frac{16}{\Lambda}\int a\left(3(\rho^2 b)^2-2(\rho^2 b)^3\right)'\,d\rho\,dt,
\label{radaction}
\eea
where the three functions  $P_{1,2,3}(\rho)$ are defined by the 
following angular integrals
\bea
P_1(\rho) &=& 
\frac{1}{\pi}\int_0^\pi
H(\rho \cos{\theta})^{3/2}
\,\sin^2\theta\,d\theta
=\frac{1}{2}+\frac{1}{8}\rho^2
  \nonumber \\
P_2(\rho) &=&  
\frac{1}{\pi}\int_0^\pi
\frac{H(\rho \cos{\theta})^{2}-1}{H(\rho \cos{\theta})^{1/2}  }
\,\sin^2\theta\,d\theta
=\frac{1}{6}\rho^2-\frac{1}{72}\rho^4+{\cal O}(\rho^6)
   \nonumber \\
P_3(\rho) &=&  
\frac{1}{\pi}\int_0^\pi
\frac{H(\rho \cos{\theta})^{2}-1}{H(\rho \cos{\theta})^{1/2}  }
\,\sin^2\theta\,\cos^2\theta\,d\theta
=\frac{1}{12}\rho^2-\frac{5}{576}\rho^4+{\cal O}(\rho^6).
\eea
The field equation for $a(\rho)$, that follows from 
the variation of (\ref{radaction}), may be integrated once to yield
\beq
\label{aprimesolution}
a' = -\frac{8\rho b^2(3-2\rho^2 b)}{\Lambda(P_1-P_2+P_3)},
\eeq
where the constant of integration has been set to zero 
in order to have a vanishing electric field at the origin $a'(0)=0$. 
Integration by parts of the Chern-Simons term in (\ref{radaction}), together
with the solution (\ref{aprimesolution}),  
produces the following energy functional, 
that depends only on the profile function 
$b(\rho),$
\bea
\label{radenergy}
\frac{E}{2\pi^2} &=& \int \Big\{
\frac{64\rho^2b^4(3-2\rho^2b)^2}{\Lambda^2(P_1-P_2+P_3)}   
+12 P_1\big((2b+\rho b')^2+4b^2(1-b\rho^2)^2\big) \nonumber\\
  &&
-8\big(6b^2P_2+2b^3\rho^2(P_2+2P_3)(b\rho^2-2)+b'(b'\rho+4b)\rho(P_2-P_3)\big)  
\Big\}\rho^3\,d\rho.
\nonumber\\
\eea
Minimization of this energy gives a second order ordinary differential
equation for $b(\rho)$ that  must be solved subject to the boundary conditions
\beq
b'(0) = 0 \quad \mbox{and} \quad 
\rho^2b\to 1 \quad \mbox{as} \quad \rho\to\infty.
\label{probc}
\eeq
Given this profile function, $a(\rho)$ can be obtained by integrating
(\ref{aprimesolution}). We shall present this numerical 
solution at the end of this
section, but first we see how the flat space self-dual instanton
approximation fits within this formalism.

In the case of large 't Hooft coupling (which is required in top-down 
approaches) the Chern-Simons term is parametrically suppressed with 
respect to the Yang-Mills term. 
The role of the Chern-Simons coupling is to provide an electric 
contribution that stabilize the soliton against the shrinking induced
by the spacetime curvature.
Large $\Lambda$ should therefore correspond to a small soliton size,
so that space is approximately flat in the soliton core.
This motivates the use of the flat space self-dual instanton to
approximate the soliton \cite{Hong:2007kx,Hata:2007mb}.

To investigate the large $\Lambda$ limit it is useful to 
first introduce the
rescaled coordinate $\rhot=\sqrt{\Lambda}\rho.$  
The boundary condition $\rho^2b\to 1$ as $\rho\to\infty,$ determines
that the appropriate associated rescaling of the profile function is 
 $\bt=b/\Lambda.$ In terms of these variables the energy (\ref{radenergy})
can be written as $E=\sum_{j=0}^\infty E_j\Lambda^{-j},$ where the first
two terms are
\be
E_0=12\pi^2 \int \Big\{\big((2\bt+\rhot \bt')^2+4\bt^2(1-\bt\rhot^2)^2\big) 
\Big\}\rhot^3\,d\rhot,
\label{e0}
\ee
\bea
\label{e1}
E_1=\frac{\pi^2}{3} \int \Big\{
4\bt^2\big(192\bt^2(2\rhot^2\bt-3)^2+\rhot^4\bt^2-2\rhot^2\bt+6\big)
+5\rhot\bt'(\rhot\bt'+4b)
\Big\}\rhot^5\,d\rhot.
\eea
Restricting to the leading order term, the energy $E_0$ is minimized
by the profile function of the flat space self-dual instanton
\be
\bt=\frac{1}{\rhot^2+\tilde\mu^2},
\label{sdprofile}
\ee
where $\tilde\mu$ is the rescaled arbitrary size of the instanton.
The leading order term in the energy is $E_0=8\pi^2$ and is
independent of the size of the instanton.

The self-dual approximation involves restricting the profile
function to the self-dual form (\ref{sdprofile}) and using the
next order term in the energy, $E_1,$ as an energy function on the
moduli space of instanton sizes. Explicitly, substituting (\ref{sdprofile})
into (\ref{e1}) and performing the integration yields
\be
E_1=2\pi^2\bigg(\frac{2}{3}\tilde\mu^2+\frac{256}{5\widetilde\mu^2}\bigg),
\ee
which is minimized when
\be
\tilde\mu={4}\bigg(\frac{3}{10}\bigg)^{1/4}.
\label{tildesize}
\ee
Returning to unscaled variables, with 
$\mu=\tilde\mu/\sqrt{\Lambda}$ the size of the instanton,
the self-dual approximation gives
\beq
E=2\pi^2\bigg(4+\frac{2}{3}\mu^2+\frac{256}{5\Lambda^2\mu^2}\bigg)
+{\cal O}\bigg(\frac{1}{\Lambda^2}\bigg),
\label{sdenergy}
\eeq
where
\be
\mu=\frac{4}{\sqrt{\Lambda}}\bigg(\frac{3}{10}\bigg)^{1/4}.
\label{size}
\ee
A similar scaling analysis of equation (\ref{aprimesolution})
shows that the leading order result for $a'$ simply corresponds to 
replacing the term $P_1-P_2+P_3$ in (\ref{aprimesolution}) by its
flat space limit $\frac{1}{2}.$ After substituting the 
self-dual approximation 
$b=1/(\rho^2+\mu^2)$ and integrating, the result is
\be
a=\frac{8(\rho^2+2\mu^2)}{\Lambda(\rho^2+\mu^2)^2}.
\label{sda0}
\ee
Note that $a(0)={16}/({\Lambda\mu^2})=\sqrt{\frac{10}{3}}$ is
independent of $\Lambda$ within this self-dual approximation.

In summary, the first term in the energy (\ref{sdenergy}) is
independent of the instanton size and is simply the 
flat space self-dual Yang-Mills result of $8\pi^2$ in our units. 
The second term is ${\cal O}(\mu^2)$ and also derives from the 
Yang-Mills functional but from the leading order correction to the 
the metric expansion around flat space.  
This gravitational contribution drives the instanton towards zero size.
The third term is ${\cal O}(1/\mu^2)$ and is the first contribution 
from the electrostatic abelian field. This term resists the shrinking of
the instanton size. These competing effects combine to produce the 
finite size (\ref{size}), which is small for large $\Lambda,$
with the energy dominated by the flat space self-dual contribution. 
The correction from the size stabilizing terms is subleading and is
 ${\cal O }(1/\Lambda)$.
\begin{figure}
\begin{center}
\includegraphics[width=7.8cm]{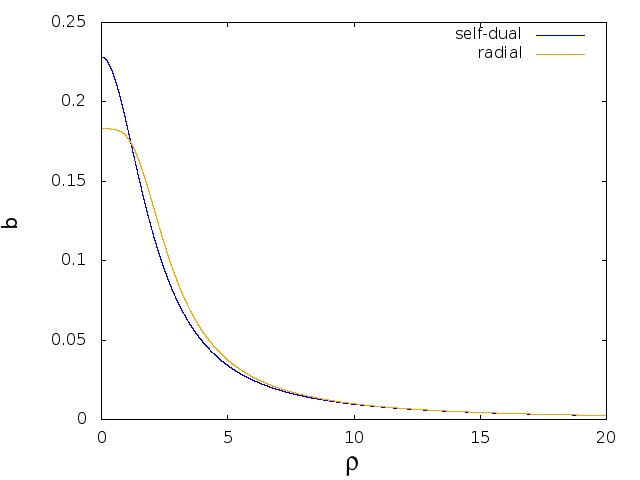}
\includegraphics[width=7.8cm]{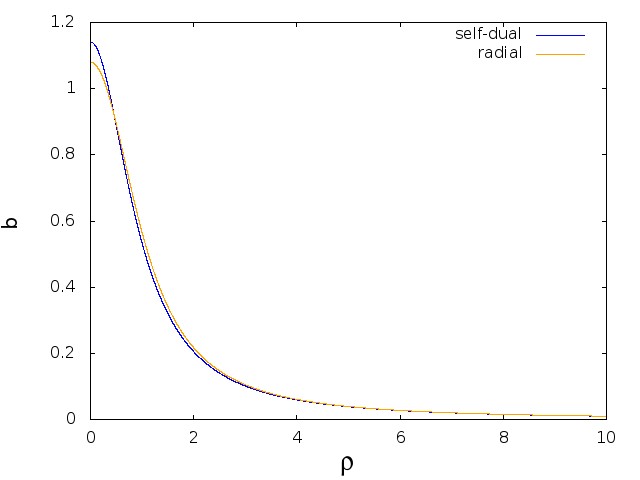}
\caption{
The profile function $b(\rho)$ 
using the flat space self-dual approximation (blue curve) 
and the radial approximation (orange curve).
The left and right images correspond to the coupling $\Lambda=2$
and $\Lambda=10$ respectively.
}\label{fig-pro} 
\end{center}
\end{figure}

Returning to the radial approximation, the profile function $b(\rho)$ that
minimizes the energy (\ref{radenergy}) subject to the boundary
conditions (\ref{probc}), was obtained using a shooting method with a 
fourth order Runge-Kutta algorithm to solve the second order ordinary 
differential equation obtained from the variation of the energy. 
The results are displayed in
Figure~\ref{fig-pro} for two values of the coupling $\Lambda=2,10.$
These plots illustrate the flow of the radial approximation to the
self-dual approximation as $\Lambda\to\infty.$
For finite $\Lambda$ the main difference between the radial and self-dual
approximations is that the self-dual approximation overestimates the value
at the origin. As we shall see later, the full numerical solution confirms
this overestimation, with the radial approximation being an 
improvement that reduces, but does not eliminate, this error.

The above rescaling to the self-dual instanton in the 
$\Lambda\to\infty$ limit is a radial restriction of the following 
rescaling used in \cite{Hong:2007kx,Hata:2007mb}
\beq
\label{scalex}
\widetilde{x}^{I} = \sqrt{\Lambda} {x}^{I}, \qquad 
\tilde t=t, \qquad
\widetilde{A}_{I} = A_{I}/\sqrt{\Lambda},  \qquad  
\widetilde{\hat{A}}_0 =\hat{A}_0.
\eeq
Defining $\tilde H=H(\widetilde{z}/\sqrt{\Lambda}),$
then in the rescaled variables the action becomes
\bea
\label{staticactionisolates}
{S }&=& 
 \int 
\left\{  -\frac{ \tilde H^{3/2}}{2 }   {\rm tr}  \left(
   \widetilde{F}_{IJ}^2\right)-\frac{ 1 - \tilde H^2}
{2  \tilde H^{1/2}}  {\rm tr}  \left(   {\widetilde{F}}_{i  j}^2\right)  \right. 
 \nonumber \\ &&  \qquad \qquad \ 
+\left. \frac{1}{\Lambda} \left( 
\frac{ \tilde H^{3/2}}{2 } (\tilde \partial_{I} \widetilde{\hat{A}}_0)^2 + 
  \frac{ 1 - \tilde H^2}{2  \tilde H^{1/2}} (\tilde \partial_i \widetilde{\hat{A}}_0)^2 \right)  \right\}
\,d^4 \widetilde{x}\, d\widetilde{z} 
  \nonumber \\
&&  + \frac{1}{\Lambda}  \int  
\ \widetilde{\hat{A}}_0 \ {\rm tr}\,( \widetilde{F}_{IJ}\widetilde{F}_{KL}) \,  
\epsilon_{IJKL} 
\,d^4 \widetilde{x}\, d\widetilde{z}. 
\eea
Using the metric (\ref{class}), with a general value of $p,$ and
expanding in $1/\Lambda$ gives
\bea
S= 
\int\bigg\{
&& -\frac{ 1}{2 }  {\rm tr} 
\left(  \widetilde{F}_{IJ}^2  \right)
\nonumber \\ 
&&+ \frac{1}{\Lambda} \left(
-\frac{3}{4} p \widetilde{z}^2 {\rm tr} \left( \widetilde{F}_{IJ}^2 \right)
+ p \widetilde{z}^2  {\rm tr} \left( \widetilde{F}_{ij}^2 \right)
+\frac{1}{2} (\tilde \partial_{I} \widetilde{\hat{A}}_0)^2
+  \widetilde{\hat{A}}_0 \ 
{\rm tr}\,( \widetilde{F}_{IJ} \widetilde{F}_{KL}) \,  \epsilon_{IJKL}   \right)
\nonumber \\
&& +{\cal O}\left(\frac{1}{\Lambda^2}\right) 
\qquad\bigg\} \,d^4 \widetilde{x}\, d\widetilde{z}, 
\eea
which highlights the convenience of the rescaling (\ref{scalex}). 
The leading order term is scale invariant and is simply 
the Yang-Mills action in flat space. 
The next term is of order $1/\Lambda$ and contains  
the size stabilizing contributions from both the abelian field
and the curvature (due to the positive value of $p$).
The action of the leading order term is minimized by the self-dual instanton
and the term of order $1/\Lambda$ defines an action on the 
self-dual instanton moduli space that fixes the size of the instanton.

In summary, the way to extract the self-dual instanton limit is to 
convert to the rescaled coordinates (\ref{scalex}) and then perform the 
$\Lambda \to \infty$ limit
\beq
\label{BPSTexpansion}
\lim_{\Lambda \to \infty} \widetilde{{\cal A}}(\widetilde{x}) = 
\widetilde{{\cal A}}_{self-dual}(\widetilde{x}),
\eeq
to converge to a self-dual instanton with a size 
$\tilde\mu$ in rescaled coordinates given by(\ref{tildesize}).
For large but finite $\Lambda$ the small unscaled instanton size is
$\mu=\tilde\mu/\sqrt{\Lambda}.$

It is important to note that the self-dual limit has nothing to say
about the asymptotic fields of the soliton at large distance.
This is because the rescaling performed in (\ref{scalex}) 
involves zooming in to a scale of order $1/\sqrt{\Lambda}$.
To study the fields of the soliton at distances greater than
$1/\sqrt{\Lambda}$ requires alternative approaches that we describe
in the next section.

\section{The soliton tail}
\label{tail}
\subsection{A linear expansion in flat space}\quad
In this subsection we consider a linear expansion that we shall
see is valid in the
region ${L}/{\sqrt{\Lambda}}\lesssim \rho \lesssim L,$ where we recall that we 
have set $L=1.$
This region is far enough from the soliton core that a linear expansion
is possible but is close enough to the origin that the curvature of the
metric can be neglected by setting $H=1.$  

To derive this expansion
we still use $1/\Lambda$ as the small parameter of the expansion, 
but now we keep the length scale fixed rather than zooming in to the core.
In this limit
\beq
\label{lambdaexpansionlimit}
\lim_{\Lambda \to \infty} \Lambda \, {\cal A}(x)= {\cal A}_{tail}(x),
\eeq
where ${\cal A}_{tail}$ is a finite term that solves the linearised 
field equations. 
The task is to compute ${\cal A}_{tail}(x)$ and to confirm its region
 of applicability.

We define the $1/\Lambda$ expansion
\bea
\label{oldlinearexpansion}
A_{I} =  A^{(1)}_{I} + A^{(2)}_{I}  + \dots, \qquad\qquad
\hat{A}_{0} = \hat{A}_{0}^{(1)} + \hat{A}_{0}^{(2)} + \dots
\eea
in which 
\beq
A_{I}^{(n)} , \,
\hat{A}_{0}^{(n)}  \propto  \frac{1}{\Lambda^n}.
\eeq
The limit (\ref{lambdaexpansionlimit}) picks up only the first term
 in this expansion
\beq
{\cal A}_{tail}(x) = \Lambda \,  {\cal A}^{(1)}(x).
\eeq
As the space is now taken to be flat, the calculation in this subsection will
involve expanding the self-dual instanton to 
provide the leading order contribution. 
This result will then be used in the next subsection to match to a
linear analysis in curved space.
  
To perform the analysis it is convenient to write the 
self-dual instanton in the gauge in which it has the 't Hooft form  
\beq
A_{I} = \frac{1}{2} 
{\sigma}_{IJ}  \partial_{J} \log{\left( 1+ \frac{\mu^2}{\rho^2}\right)}.
\eeq
Given that $\mu^2={\cal O}(1/\Lambda)$ then 
the first term in the expansion is
\beq
\label{firstA}
A^{(1)}_{I} = - {\sigma}_{IJ} \frac{x_{J} \, \mu^2}{\rho^4} = 
\frac{\mu^2}{2} {\sigma}_{IJ}  \partial_J \frac{1}{\rho^2} 
\propto \frac{1}{\Lambda},
\eeq
which satisfies the field equations 
((\ref{feq1}) and (\ref{feq2}) with $H=1$) at the linear level
since
\beq
\partial_{I}A^{(1)}_{I} = 0 \qquad \mbox{and} \qquad
 \partial_{J}\partial_J A^{(1)}_{I} =0.
\eeq
These equations are simply those of an abelian gauge potential:
the first is the condition of Coulomb gauge and the second 
is the vanishing of the Laplacian. 

The term $A^{(1)}_{I}$ gives the dominant contribution to the field 
strength 
\bea
F_{IJ}^{(1)} &=& \partial_{I}A^{(1)}_{J}- \partial_{J}A^{(1)}_{I} 
=  \frac{2 \mu^2}{\rho^4} \left(  {\sigma}_{IJ}   + 
\frac{2}{\rho^2}( {\sigma}_{JK} {x}_{K} {x}_{I}  
- {\sigma}_{IK} {x}_{K} {x}_{J})   \right).
\label{fij1}
\eea
From (\ref{sda0}) the abelian gauge potential at linear order is 
\beq
\label{firsta}
\hat{A}_{0}^{(1)} = \frac{8}{ \Lambda \rho^2}, 
\eeq
which satisfies the final field equation ((\ref{feq3}) with $H=1$) 
at linear order.

Defining
$F_{IJ}^{(2)} = \partial_{I}A^{(2)}_{J}- \partial_{J}A^{(2)}_{I},$
at second order the field equations are
\bea
\partial_I   {F}^{(2)}_{IJ} +  i [A_I^{(1)}, {F}^{(1)}_{IJ}]   =   0  
\qquad \mbox{and} \qquad
\partial_{I} \partial_{I} \hat{A}_{0}^{(2)}=0,   
\label{firstequationsecondorder}
\eea
with solution
\be
A^{(2)}_I =   {\sigma}_{IJ} \frac{x_{J} \, \mu^4}{\rho^6}  
\propto \frac{1}{\Lambda^2}, \qquad\qquad
\hat A_0^{(2)}=0.
\ee 
The next non-zero term in $\hat A_0$ is at third order, where the field equation
gives
\be
\partial_{I} \partial_{I} \hat{A}_{0}^{(3)}=\frac{1}{\Lambda} {\rm tr}\,( {F}^{(1)}_{IJ}{ F}^{(1)}_{KL}) \,  \epsilon_{IJKL},
\ee
and is solved by
\be
\hat{A}_{0}^{(3)} =
 - \frac{8 \mu^4}{ \Lambda \rho^6} \propto \frac{1}{\Lambda^3}.
\ee
For this expansion to be reliable requires $||A^{(1)}_{I}|| \lesssim ||A^{(2)}_{I}||$ 
and $||\hat{A}_{0}^{(1)}|| \lesssim ||\hat{A}_{0}^{(3)}||.$ 
These conditions correspond to the requirement that 
$\rho \gtrsim 1/\sqrt{\Lambda},$ which means far from the soliton core.
The use of the flat space metric approximation, $H=1,$ 
required $\rho\lesssim 1$, so combining these constraints results in the 
region of validity  
$\frac{1}{\sqrt{\Lambda}}\lesssim \rho \lesssim 1,$ as claimed at the start of
this subsection. 

\subsection{A linear expansion in curved space}\quad
We now extend the linear expansion of the previous subsection 
to distances beyond the restriction $\rho \lesssim 1.$ 
This requires that the curvature 
of the metric is now taken into account and the approximation $H=1$
can no longer be used. The linear analysis in this subsection is
equivalent to that in \cite{Hashimoto:2008zw} and produces the same
result. However, the derivation is a little different as we wish to
elucidate the aspects that will play a role in our additional analysis
later in the paper.

For the purposes of this subsection it will be sufficient to consider
only the first order terms $A^{(1)}_{i}$ and $\hat{A}^{(1)}_0$.
As these terms satisfy the linearised field equations we can perform a 
separation of variables in $x_i$ and $z$, expand in eigenfunctions of the 
linear operator in flat space, and then extend each eigenfunction 
separately into the curved region beyond $\rho \lesssim 1$ .  
The existence of an overlap region 
$\frac{1}{\sqrt{\Lambda}}\lesssim \rho \lesssim 1,$ in which the linear flat
space approximation and the linear curved space approximation
are both valid, allows the computation of the coefficients of the eigenfunction
expansion in curved space.

The easiest case is that of the abelian potential $\hat{A}_0^{(1)},$
which satisfies the linearized field equation (\ref{feq3}) given by
\be
\partial_i\partial_i\hat A_0^{(1)}
+H^{1/2}\partial_z(H^{3/2}\partial_z\hat A_0^{(1)})=0.
\ee
We can therefore extend (\ref{firsta}) to the curved regime by writing
\bea
\label{generalfora}
\hat{A}_{0}^{(1)} &=&  \frac{8}{ \Lambda}  \xi(x_{I})
\eea
where $\xi(x_{I})$ is a harmonic function in the four-dimensional
curved space, which in the flat regime is
\beq
\label{almostflatf}
\xi(x_{i},z) \simeq \frac{1}{\rho^2} \qquad {\rm for} \quad \rho \lesssim 1.  
\eeq
We now separate variables $x_{I} = (x_{i},z)$ and write 
$r=\sqrt{x_1^2+x_2^2+x_3^2}$ for the three-dimensional radius.
The harmonic function can be expanded in a Laplace-Fourier expansion 
(Laplace expansion in $r$, Fourier expansion in $z$).
In flat space there is the exact identity
\beq
\label{laplacefourier}
\frac{1}{\rho^2}   = \frac{1}{r^2 + z^2}  =    \int_0^{\infty}  \frac{e^{-k r}}{r} \cos{(k z)}\,dk.  
\eeq
Note that {\it all} the momentum modes $k$ must appear in this
 expansion in order to reconstruct the function $1/\rho^2$ exactly.
Now we extend this expansion into the curved region by replacing it with 
\beq
\label{newf}
\xi(x_{i},z) =   \int_0^{\infty}   \frac{e^{-k r}}{r} \psi_{(k)}^{+}( z)
\,dk,
\eeq
where $ \psi_{(k)}^{\pm}(z)$ are defined as the eigenfunctions 
satisfying the linear equation
\beq
\label{eqpsi}
H^{1/2}\partial_z (H^{3/2} \partial_z \psi_{(k)}^{\pm}) + k^2 \psi_{(k)}^{\pm} =0,
\eeq
with the superscript $^\pm$ referring to  even and odd parity  with respect to $z \to -z$.
The  boundary conditions for $\psi_{(k)}^{+}(z)$ are
\beq
\label{boundaryconditionpsi}
 \psi_{(k)}^{+}(0) = 1, \qquad  \qquad \partial_z \psi_{(k)}^{+}(0) = 0.
\eeq
Only the even eigenfunctions $\psi_{(k)}^{+}(z)$  appear in the expansion for 
$\xi(x_I)$, but later we shall need the odd eigenfunctions $\psi_{(k)}^{-}(z),$ 
which satisfy the boundary conditions  
\beq
\label{boundarypsiminus}
 \psi_{(k)}^{-}(0) = 0, \qquad  \qquad \partial_z \psi_{(k)}^{-}(0) = 1.
\eeq

The expression (\ref{generalfora}) with 
$\xi(x_{I})$ defined in (\ref{newf}) gives the {exact} 
 extension of $\hat{A}_0^{(1)}$ in the curved region and reduces to 
(\ref{almostflatf}) in the almost flat region since,
for every value of $k$,
\beq
\psi_{(k)}^{+}(z) \simeq  \cos{(k z)}  \qquad {\rm for}  \qquad z \ll 1,
\eeq
as $H\simeq 1$ in this region.

Next we consider the non-abelian field $A_{I}^{(1)},$ given by (\ref{firstA})
in the flat regime. First we decompose into parity components
\bea
\label{generalforAintermediate}
A^{(1)}_{i} = A^{(1+)}_{i} + A^{(1-)}_{i}, \qquad\qquad
A^{(1)}_{z} = A^{(1+)}_{z}
\eea
where the superscript $^\pm$ again stands for the parity with respect 
to $z \to -z$.
The odd component $A^{(1-)}_{z}$ vanishes in the chosen gauge where
$\partial_iA_i^{(1+)}=0.$

In the flat regime (\ref{firstA}) gives the parity components 
\be
A^{(1+)}_{i} = \frac{\mu^2}{2} \epsilon_{ijk}{\sigma}_{k} \partial_j \frac{1}{\rho^2}, \qquad  
A^{(1-)}_{i} =  - \frac{\mu^2}{2} {\sigma}_{i} \partial_{z} \frac{1}{\rho^2}, \qquad
A^{(1+)}_{z}  = \frac{\mu^2}{2} {\sigma}_{i}  \partial_i\frac{1}{\rho^2}.
\label{blocks}
\ee
Applying the parity decomposition to the linearized field equations
(\ref{feq1}) and (\ref{feq2}) yields
\bea
&&\partial_j\partial_j A_i^{(1+)}
+H^{1/2}\partial_z(H^{3/2}\partial_z A_i^{(1+)})=0,
\label{lfeq1}\\
&&\partial_i(\partial_i  A_z^{(1+)}-\partial_z A_i^{(1-)})=0,
\label{lfeq2}\\
&&\partial_j(\partial_j A_i^{(1-)}-\partial_i A_j^{(1-)})
+H^{1/2}\partial_z(H^{3/2}\partial_z A_i^{(1-)})
-\partial_i(H^{1/2}\partial_z(H^{3/2} A_z^{(1+)}))=0.\qquad 
\label{lfeq3}
\eea
The easiest component to deal with is $A_i^{(1+)}$ as this decouples
from the other components and satisfies the same equation as
the abelian potential $A_0^{(1)}.$
The first component in (\ref{blocks}) is therefore extended to
curved space as 
\bea
\label{generalforAiplus}
A^{(1+)}_{i} &=& \frac{\mu^2}{2}  \epsilon_{ijk}{\sigma}_{k} \partial_{j} \xi(x_{I}), 
\eea
where $\xi(x_{I})$ is the same harmonic function defined in (\ref{newf}). 
Note that here, as for $\hat{A}_0$, only the even eigenfunctions 
appear in the expansion.

The remaining two components $A_i^{(1-)}$ and $A_z^{(1+)}$ 
are slightly more complicate as their equations (\ref{lfeq2}) and
(\ref{lfeq3}) are coupled together.
We see from (\ref{lfeq2}) that if  $A_i^{(1-)}$ is expanded using the
eigenfunctions $\psi_{(k)}^-$ 
then $A_z^{(1+)}$ must be expanded in terms
of their derivatives $\partial_z\psi_{(k)}^-,$
which then gives a consistent expansion for (\ref{lfeq3}). 
We therefore define the functions
\beq
\label{phidpsi}
\phi_{(k)}^{\pm}(z) =  \partial_z \psi_{(k)}^{\mp}( z).
\eeq
In the almost flat region $H\simeq 1,$ so we have that
\bea
 \psi^{-}_{(k)}(z) \simeq  \frac{\sin{(kz)}}{k} \quad {\rm and } \quad \phi_{(k)}^{+}(z) \simeq  \cos{(k z)}  \qquad {\rm for}  \qquad z \ll 1.
\eea
The flat space results (\ref{blocks}) may be rewritten as
\bea
A^{(1-)}_{i} = \frac{\mu^2}{2} {\sigma}_{i}  
\int_0^\infty 
\frac{e^{-kr}}{r} k \sin{(kz)}\,dk, \qquad
A^{(1+)}_{z}  =\frac{\mu^2}{2}  
{\sigma}_{i}  \int_0^\infty  \,\partial_i \frac{e^{-kr}}{r} \cos{(kz)}\,dk,
\eea
so the extension to curved space is
 \bea
\label{generalforAizminus}
A^{(1-)}_{i} =  \frac{\mu^2}{2} {\sigma}_{i} \int_0^\infty \frac{e^{-kr}}{r} k^2  \psi^{-}_{(k)}(z)\,dk, \qquad 
A^{(1+)}_{z}  = \frac{\mu^2}{2} {\sigma}_{i} \int_0^\infty   \partial_i \frac{e^{-kr}}{r} \phi^{+}_{(k)}{(z)}\,dk,
\eea
where only the odd eigenfunctions $\psi_{(k)}^-$ and their 
derivatives $\phi_{(k)}^{+}$ appear in the expansions of these components.
As for the other components described earlier, 
the expressions (\ref{generalforAizminus}) give the {exact}
 extensions of $A_i^{(1-)}$ and $A_z^{(1+)}$ to  the  curved  region,
 and  reduce to (\ref{blocks}) in the almost flat region.

\subsection{The effect of a conformal boundary}\quad
The expressions (\ref{generalfora}), (\ref{generalforAiplus}) and 
(\ref{generalforAizminus}) are  exact identities, but only if all the 
momentum modes $k$  are taken into account.
A case in which this is compatible with the boundary conditions is when 
the metric does not have a conformal boundary, for example if $p<\frac{1}{2}.$
In this case the formulae (\ref{generalfora}), 
(\ref{generalforAiplus}) and (\ref{generalforAizminus}) 
provide the exact solution to the first order term in the linear expansion. 
Since the boundary is at conformal infinity this is the end of the story.

In contrast, if there is a conformal boundary, as in all cases in which 
an AdS/CFT interpretation is possible (including the \ss model),  
the boundary conditions for the fields at the conformal boundary 
must be specified, and this may restrict the allowed momenta $k$ in the 
Laplace-Fourier expansion.
For holographic QCD, the correct holographic prescription at the
boundary is that there are no sources for the operators in the dual theory. 
In conformal coordinates this corresponds to the field strength having 
vanishing parallel components at the boundary $z=\pm\infty$ 
for both the abelian and non-abelian fields.
In terms of the eigenfunction expansion, this condition 
translates to the boundary condition on the even eigenfunctions
\beq\label{conditionpsi1} \psi_{(k)}^{+}(\infty ) =  0, \eeq
which selects only a discrete set of momenta, 
$k_{2n-1} $ with $n= 1,2, \dots$ and  $k_{1} > 0.$ 
Similarly, the odd eigenfunctions are required to satisfy
\beq\label{conditionpsi2} \psi_{(k)}^{-}(\infty ) =  0, \eeq
which selects the discrete momenta $k_{2n} $ with $n= 1,2, \dots.$ 
It can be shown that the even and odd momenta interlace so that
we may impose the ordering $k_{n+1} > k_{n}$.

\begin{figure}
\begin{center}
\includegraphics[width=10cm]{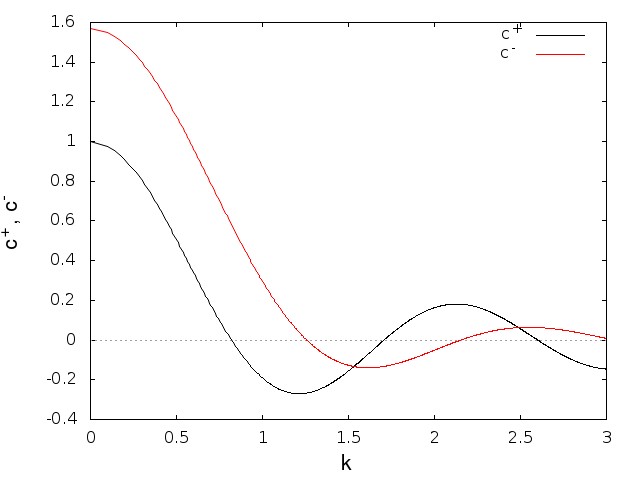}
\caption{The limit values $c^\pm(k),$ with zeros corresponding to the
allowed momenta. 
}\label{fig-evalues} 
\end{center}
\end{figure}

Restricting to the \ss value $p=\frac{2}{3},$ we may define
$c^\pm(k)=\lim_{z\to\infty}\psi_{(k)}^\pm(z),$ so that the allowed values
of the momenta $k$ are given by the zeros of $c^\pm(k).$
In Figure~\ref{fig-evalues} we plot the limit values $c^\pm(k)$
for $0\le k\le 3.$ This allows the first few values of the discrete
momenta to be determined and in particular $k_1=0.82$ and $k_2=1.26,$
which agrees with the results in \cite{Sakai:2004cn}.

As the odd (even) values of $n$ correspond to even (odd) 
functions with respect to $z \to -z$, a more convenient notation from
now on is to label the eigenfunctions by an integer by defining
\beq
\psi_{2n - 1}(z) \equiv \psi_{(k_{2n-1})}^{+}(z)\ , \qquad 
\psi_{2n}(z) \equiv \psi_{(k_{2n})}^{-}(z)\ , \qquad n=1,2,\dots
\eeq
so that the information about the parity of the eigenfunction 
is encoded in the parity of the integer index. 

In the absence of a conformal boundary the eigenfunction $\psi_{(k)}^{\pm}(z)$ 
has an infinite number of zeros and oscillates as $z \to \infty.$  
In this situation a function like $1/\rho^2$, 
that vanishes as $z \to \infty$, can be 
expanded as an integral over all momenta $k,$ as in 
(\ref{laplacefourier}), even if 
$\psi_{(k)}^{\pm}(z)\centernot\to 0$ as $z\to\infty$ 
It is the oscillating property of the eigenfunction that produces 
decoherence and leads to this result.

In contrast, when there is a conformal boundary 
the  eigenfunction  $\psi_{(k)}^{\pm}(z)$ has a finite number of zeros
and does not oscillate for large $z,$ but rather tends monotonically 
to a finite limit as $z \to \infty$. For $p=\frac{2}{3}$ the large
$z$ behaviour is $\psi_{(k)}^{\pm}(z)\sim c^\pm(k)+d^\pm(k)/z+\ldots.$ 
In this situation, the expansion of a function that vanishes at infinity
requires all the eigenfunctions in the expansion to also vanish at infinity,
so the boundary conditions (\ref{conditionpsi1}) and (\ref{conditionpsi2})
must be imposed.

To make sense of (\ref{newf}) we must therefore project to the subspace of 
allowed eigenfunctions to obtain the form
\beq
\label{newdiscretef}
\xi(x_{i},z) =  \sum_{n=1}^{\infty}   \, \xi_{2n-1} \frac{e^{-k_{2n-1} r}}{r} 
\psi_{2n-1}( z),
\eeq
where the projection coefficients $\xi_{2n-1}$ are defined by
\be
\xi_{2n-1}=\frac{1}{(\psi_{2n-1},\psi_{2n-1})}\int_0^\infty 
(\psi^+_{(k)},\psi_{2n-1})\,dk,
\label{projcoeffs1}
\ee
using the inner product 
\be
(\psi,\tilde\psi)=\int_{-\infty}^\infty \frac{1}{H^{1/2}}
\psi\tilde\psi\,dz.
\label{inprod1}
\ee
This is the inner product in which the eigenfunctions
$\psi_n$ are orthogonal, $(\psi_m,\psi_n)\propto\delta_{mn}.$

The discretization (\ref{newdiscretef}) has an important consequence. 
As $k_1>0$, 
the large distance decay is now exponential not algebraic.  
Since $\hat{A}_0$ is the field dual to the baryon current of the 
boundary theory, this means that the baryon form factors, at least within
this linear approximation, decay exponentially. 

Another consequence of the discretization appears when we 
retract back to the flat regime, 
as it converts the identity (\ref{laplacefourier}) into the
approximation 
\be
\frac{1}{\rho^2}\simeq 
\sum_{n=1}^{\infty}   \, \xi_{2n-1} \frac{e^{-k_{2n-1} r}}{r} \cos(k_{2n-1}z),
\label{projcoeffs}
\ee
due to the projection to an incomplete subset of eigenfunctions. 
However, in the almost flat region  $|z| \ll 1$ the large momenta modes are
the most important in reconstructing the function 
${1}/{\rho^2},$ so this discretization 
does not affect the validity of the small instanton approximation. 

A similar story applies to the projection of the non-abelian potential.
In particular, the relations (\ref{generalforAizminus}) become
 \bea
\label{newdiscreteAminus}
A^{(1-)}_{i} =  \frac{\mu^2}{2} {\sigma}_{i} \sum_{n=0}^{\infty} \xi_{2n} 
\frac{e^{-k_{2n}r}}{r} k_{2n}^2  \psi_{2n}(z),\qquad
A^{(1+)}_{z}  = \frac{\mu^2}{2} {\sigma}_{i}   \sum_{n=0}^{\infty} \xi_{2n} 
\partial_i \frac{e^{-k_{2n}r}}{r} \phi_{2n}{(z)},
\eea 
where $\phi_{n}(z) =  \partial_z \psi_{n}(z)$ using our new notation.
An expression for the projection coefficients $\xi_{2n}$ will be given
below, but first we draw attention to an important point.
An additional mode has been included in the expansions 
(\ref{newdiscreteAminus}), where we have defined $k_0=0.$
The associated zero mode is  
\beq
\psi_{0}( z) = \int_0^{z}   \frac{1}{H(z)^{3/2}}\,dz 
\quad\mbox{with}\quad
\phi_{0}(z) = \frac{1}{H(z)^{3/2}}.
\eeq
Note that $\psi_0(\infty)\ne 0,$ hence this mode was excluded from the
earlier considerations. However, since $k_{0}=0$ this mode does not
contribute to $A_{i}^{(1-)}$  due to the $k_{2n}^2$ factor 
in the first formula in (\ref{newdiscreteAminus}).
Thus the boundary condition on the field strength (that the parallel 
components vanish at the conformal boundary) remains satisfied.
The eigenfunction $\psi_0$, with zero eigenvalue, is associated with the
massless pion and contributes through the inclusion in $A_z^{(1+)}$
of the mode $\phi_0,$ which vanishes at infinity.
For the \ss model $p=\frac{2}{3}$ which gives $\phi_0=1/(1+z^2)$ and
$\psi_0=\tan^{-1}z$ with $\psi_0(\infty)=\frac{\pi}{2}.$

From the second formula in (\ref{newdiscreteAminus}) the projection
coefficients are given by a similar expression to (\ref{projcoeffs1}),
namely
\be
\xi_{2n}=\frac{1}{\langle\phi_{2n},\phi_{2n}\rangle}\int_0^\infty 
\langle\phi^+_{(k)},\phi_{2n}\rangle\,dk,
\label{projcoeffs2}
\ee
using the appropriate inner product 
\be
\langle\phi,\tilde\phi\rangle=\int_{-\infty}^\infty {H^{3/2}}
\phi\tilde\phi\,dz,
\label{inprod2}
\ee
for orthogonality $\langle\phi_m,\phi_n\rangle\propto\delta_{mn}.$

For $n\ne 0,$ an integration by parts, 
together with an application of the defining equation for the eigenfunctions 
(\ref{eqpsi}), proves the identity
\be
\langle\phi_{(k)}^+,\phi_{2n}\rangle=(k^2\psi_{(k)}^-,\psi_{2n}).
\ee
Using this identity gives the first projection formula
in (\ref{newdiscreteAminus}), which completes the derivation.

In summary, the conclusion from the linear analysis in curved space is
that at large three-dimensional distance, $r \gtrsim 1$,
 all terms decay exponentially,
except the algebraic decay associated with the pion field. 
Explicitly,
\bea
A^{(1)}_{z}  = -\frac{\xi_0 \mu^2}{2} \, \frac{\sigma_{i}\hat{x}_i}{r^2} \, \phi_{0}( z) + {\cal O} \bigg(\frac{e^{ -k_{2}r}}{r}\bigg), \quad
A^{(1)}_{i} =  {\cal O} \bigg(\frac{e^{ -k_{1}r}}{r}\bigg), \quad
\hat A^{(1)}_{0} =  {\cal O} \bigg(\frac{e^{ -k_{1}r}}{r}\bigg),
\label{lincurved}
\eea
where $\hat x_i=x_i/r.$
In the following subsection we shall show that the linear result
(\ref{lincurved}) cannot be used to conclude anything about the 
asymptotic tail of the soliton fields. In particular, by extending it
to arbitrarily large values of the radius $r,$ it leads to incorrect
conclusions regarding the exponential decay of physical quantities
of the baryon, such as the baryon density and 
electromagnetic form factors.
We shall see that these linear results do have a region of validity, but
this region does not include arbitrarily large values of the radius, since
nonlinear terms then dominant over the linear result (\ref{lincurved}).
This is the source of several erroneous computations and conclusions in the
literature. 

\subsection{Noncommutativity of the large $\Lambda$ and large $r$ limits}\quad 
As we saw in the previous subsection, 
if we take the leading order term in the $1/\Lambda$ expansion
and then expand again to find the large $r$ behaviour, then the  
dominant contribution is from the pion field. 
It produces an ${\cal O}(1/r^2)$ term that appears {\it only} in the 
$A_z$ component of the gauge potential
 and not in the $A_i$ or $\hat{A}_0$ components, which
decay exponentially. 
It is crucial to note the order of the limits here: 
first we take the large $\Lambda$ limit,
which selects the linear term ${\cal A}^{(1)},$ and then 
we consider the large $r$ limit.
This ordering assumes that the linearized fields given in (\ref{lincurved}) 
provide the dominant contribution at large $r.$ 
If this assumption is to be valid then it requires that 
all higher order terms in the $1/\Lambda$ expansion of
$A_i$ and $\hat{A}_0$ decay exponentially with $r.$
In this subsection we prove that this 
requirement is not satisfied and hence the linear 
result is not valid at large $r.$ 
We begin by assuming the linear result is valid and then find a 
contradiction. 

For the remainder of the computations in this subsection we ignore all
terms that decay exponentially with $r,$ as we are interested in the
details of the algebraic decay. 
With the exponential terms neglected, $A_i=\hat A_0=0.$
From (\ref{lincurved}), the only non-zero components of the 
field strength at linear order in $1/\Lambda$ are
 \beq
\label{fieldlinearpion}
F^{(1)}_{iz} =\frac{\xi_{0} \mu^2}{2} 
\phi_0 \sigma_j \partial_i\partial_j \bigg(\frac{1}{r}\bigg). 
\eeq
In particular, this means that at the linear level the
instanton charge density decays exponentially.

The field equations (\ref{feq1}) and (\ref{feq2}) now become
\beq
\label{eqmot}
D_z \left(H^{3/2}F_{iz}\right)=0, \qquad\qquad
 D_i F_{iz}  =0.
\eeq
We can check that they are satisfied at linear order
\bea
\label{sollin}
\partial_z \left(H^{3/2} F^{(1)}_{iz}\right) =  0, \qquad\qquad 
\partial_i F^{(1)}_{iz} = 0,  
\eea
using (\ref{fieldlinearpion}) and the identities
\be
\partial_z \left(H^{3/2} \phi_{0}\right) = 0, \qquad\qquad
\partial_i \partial_i \left( \frac{1}{r} \right) =  0.
\ee
However, a problem arises at the next order in the $1/\Lambda$ expansion.
At second order the first equation in (\ref{eqmot}) becomes
\be
\partial_z(H^{3/2}\partial_i A_z^{(2)})+iH^{3/2}[A_z^{(1)},F^{(1)}_{iz}]=0,
\ee
which simplifies to
\be
\partial_z(H^{3/2}\partial_i A_z^{(2)})=-\frac{1}{2}\xi_0^2\mu^4\phi_0
\epsilon_{ijk}\frac{\hat x_j}{r^5}\sigma_k.
\ee
This equation determines the $z$ dependence of $A_z^{(2)}$ to be
\be
A_z^{(2)}=-\frac{1}{2}\xi_0^2\mu^4\phi_0\psi_0\beta,
\ee
where $\beta$ is independent of $z$ and solves the equation
\be
\partial_i\beta=\epsilon_{ijk}\frac{\hat x_j}{r^5}\sigma_k.
\label{betaeqn}
\ee
However, it is easy to prove that there are no solutions to (\ref{betaeqn}).
Defining the right hand side of (\ref{betaeqn}) to be $\Xi_i,$ the
existence of a solution $\beta$ requires the zero curvature
condition $\partial_i\Xi_j-\partial_j\Xi_i=0,$ which is easily calculated
and does not vanish.
    
This proves that it is impossible to extend the linear result
(\ref{lincurved}) to higher order in $1/\Lambda$ if 
$A_i$ and $\hat A_0$ decay exponentially with $r.$
Terms in $A_i$ and $\hat A_0$ with algebraic decay are required beyond
linear order for a consistent expansion.
As a result, at large radius these higher order terms in $1/\Lambda$ 
dominate over the exponential terms at linear order.
The upshot is that the linear result is not valid at large radius
and gives incorrect results for physical quantities, 
such as the baryon density, the abelian electric field
and electromagnetic form factors.
In the following subsection we derive the correct extension of
the linear expansion (\ref{lincurved}) for large $r.$

\subsection{A nonlinear expansion at large $r$}\quad
\label{newlinear}
The $1/\Lambda$ expansion will still play a role in this subsection, but
to obtain the correct large $r$ behaviour it is vital to include 
higher order terms beyond the linear contribution ${\cal A}^{(1)}.$ 
We reverse the order of the limits in the previous subsection by
first considering the large $r$ limit and then performing the 
$1/\Lambda$ expansion.
Explicitly, we keep the leading order terms in a $1/r$ expansion
at each order in a $1/\Lambda$ expansion.
As in the previous subsection, we ignore all exponentially decaying terms,
so from (\ref{lincurved}) the expansion starts with the linear term 
in $1/\Lambda$ 
\beq
\label{newlinearstart}
A^{(1)}_{z}  = -\frac{\xi_{0} \mu^2}{2} \, 
\frac{\sigma_{i}\hat{x}_i}{r^2} \, \phi_{0}, 
\qquad A^{(1)}_{i}  =0, 
\qquad \hat A^{(1)}_0  =0.
\eeq
As confirmed previously, the field equations are satisfied
at linear order. 
At second order the field equation (\ref{feq1}) becomes
\bea
\label{fullequation}
\partial_z(H^{3/2} F^{(2)}_{zi}) + i H^{3/2}   [A^{(1)}_{z}, F^{(1)}_{zi}] 
 +H^{-1/2} \partial_j F^{(2)}_{ji} = 0. 
\eea
At large $r$ the final term in this expression is of lower order
in a $1/r$ expansion than the first and may be neglected.
This leaves
\be
\partial_z(H^{3/2}(\partial_iA^{(2)}_z-\partial_zA^{(2)}_i))
 + i H^{3/2}   [A^{(1)}_{z}, F^{(1)}_{iz}] =0.
\label{seccurved}
\ee
As we saw in the previous subsection, it is impossible to solve
this equation with $A^{(2)}_i=0.$ 
We now derive the solution for $A^{(2)}_i$ in the gauge $A_z^{(2)}=0,$ 
when (\ref{seccurved}) becomes
\beq
\label{excesscancelled}
\partial_z(H^{3/2}\partial_z A_i^{(2)})=
\frac{1}{2}\xi_0^2\mu^4\phi_0 \epsilon_{ijk}\frac{\hat x_j}{r^5}\sigma_k.
\eeq
The $z$ dependence factors as 
\beq
\label{secondorderterm}
 A_{i}^{(2)} = \eta \frac{1}{2}\xi_0^2\mu^4\epsilon_{ijk}\frac{\hat x_j}{r^5}\sigma_k,
\eeq
where $\eta(z)$ solves the equation
\beq
\partial_z(H^{3/2}\partial_z \eta)=\phi_0.
\eeq
Using $\phi_0=1/H^{3/2}=\partial_z\psi_0,$ this equation may
be integrated once to give
\be
\partial_z \eta = \psi_{0}\phi_{0},
\ee
which is solved by
\beq
\label{eqeta}
\eta = \frac{1}{2}\psi_{0}^2 - \frac{\pi^2}{8},
\eeq
where the constant of integration has been fixed by the requirement that
 $\eta(\infty)=0$ and we have used the earlier result that
$\psi_0(\infty)=\frac{\pi}{2}$ for the \ss model with $p=\frac{2}{3}.$

The $ij$ component of the field strength has a contribution at second order 
\beq
\label{fieldsecondnew}
 F_{ij}^{(2)} =-\frac{\xi_0^2\mu^4\eta}{r^6} 
\big(\epsilon_{ijk} + 3\hat x_l(\epsilon_{ikl}\hat x_j-\epsilon_{jkl}\hat x_i)\big)\sigma_k 
\eeq
and thus the term proportional to the instanton charge density
\be
{\cal I}=   \epsilon_{IJKL} \,  {\rm tr}\,( {F}_{IJ}{ F}_{KL})   
\ee
is generated at third order in $1/\Lambda$
\be
   {\cal I}^{(3)}=
4\epsilon_{ijk} \,  {\rm tr}\left(F^{(2)}_{ij}F^{(1)}_{kz}\right)
=-\frac{48 \xi_0^3 \mu^6 \phi_0\eta}{r^9} \propto  \frac{1}{\Lambda^3 r^9}.
\ee

The abelian field $\hat{A}_0$ is sourced by ${\cal I}$ with a 
coupling $1/\Lambda$, and thus it is generated at fourth order. 
The equation for $\hat{A}_0$ in radial coordinates is
\bea
H^{-1/2}\partial_{r} \left(r^2 \partial_{r} \hat{A}_0\right) 
+ r^2 \partial_z \left(H^{3/2} \partial_z \hat{A}_0\right) = 
\frac{1}{\Lambda}  r^2  {\cal I},
\eea
so the fourth order term satisfies
\bea
H^{-1/2}\partial_{r} \left(r^2 \partial_{r} \hat{A}_0^{(4)}\right) 
+ r^2 \partial_z \left(H^{3/2} \partial_z \hat{A}_0^{(4)}\right) = 
-\frac{48 \xi_0^3 \mu^6 \phi_0\eta}{\Lambda r^7}.
\eea
Applying the ansatz
\beq
\label{azerofour}
\hat{A}_0^{(4)} = -\frac{48 \xi_0^3 \mu^6}{\Lambda r^9}\chi,
\eeq
with $\chi(z)$, and neglecting subleading terms in $1/r$, 
we obtain the equation
\beq
\label{eqvarrho}
 \partial_z \left(H^{3/2} \partial_z \chi \right) = \phi_0\eta,
\eeq
which must be solved subject to the boundary conditions 
$\chi(\pm \infty) = 0$.
The solution is easily obtained by using $\psi_0$ as the 
independent coordinate rather than $z$, as equation (\ref{eqvarrho}) 
then simplifies to 
\beq
\frac{\partial^2 \chi}{\partial \psi_0^2} = 
\frac{1}{2}\psi_{0}^2- \frac{\pi^2}{8}.
\eeq
The unique solution satisfying the above boundary conditions
is
\beq
\chi=  
  \frac{1}{24}  \bigg(\psi_{0}^4  -6 \bigg(\frac{\pi}{2}\bigg)^2 \psi_{0}^2 
+5\bigg(\frac{\pi}{2}\bigg)^4\bigg),
\eeq
to give
\be
\hat{A}_0^{(4)} = -\frac{2 \xi_0^3 \mu^6}{\Lambda r^9}
\bigg(\psi_{0}^4  -6 \bigg(\frac{\pi}{2}\bigg)^2 \psi_{0}^2 
+5\bigg(\frac{\pi}{2}\bigg)^4\bigg).
\label{azerofour2}
\ee
We have now achieved our aim of determining the leading order large
$r$ behaviour of all the fields and their relation to the small
instanton approximation. Namely,
\be
A_{z}  = -\frac{1}{2}\xi_{0} \mu^2  
\frac{\hat{x}_i\sigma_i}{r^2}  \phi_{0}
+\ldots, 
\qquad 
A_{i} = \frac{1}{2}\xi_0^2\mu^4\epsilon_{ijk}\frac{\hat x_j\sigma_k}{r^5}
\eta +\ldots 
, 
\qquad 
\hat A_0  =-\frac{48 \xi_0^3 \mu^6}{\Lambda r^9}\chi
+\ldots
\ee
Note the significant difference in the rate of decay of the
abelian field $\hat A_0$ in the $r$ and $z$ directions, since $\chi$ 
decays only as ${\cal O}(\frac{1}{z})$ for large $z.$

\subsection{The emergence of a new scale}\quad
We now describe the way in which the results we have obtained
imply the existence of a new large scale, 
in which the behaviour of the $A_i$ and $\hat{A}_0$ components
are dominated by nonlinear terms.
Recall that the $1/\Lambda$  expansion at large $r$ takes the form
\bea
A_{z} &=&  A^{(1)}_{z} + \dots \nonumber \\
A_{i} &=& \phantom{ A^{(2)}_{z} +}  A^{(2)}_{i} + \dots \nonumber \\
\hat{A}_{0} &=&  \phantom{ A^{(1)}_{z} +  A^{(3)}_{i} + \ \ \ \ \ } \hat{A}_0^{(4)} + \dots \label{nontail}
\eea
where $A_{i}$ starts only at second order and 
$\hat{A}_0$ starts only at fourth order, once exponentially decaying
terms are neglected. However, we need to determine the scale at which
it is appropriate to neglect these exponentially decaying terms, so that
the nonlinear terms with algebraic decay dominate over the 
linear result.

The new scale is where the linear terms in the $1/\Lambda$ expansion
of $A_i$ and $\hat A_0$  are comparable to the higher order terms, that is,
$A^{(1)}_i \sim A^{(2)}_i,\ \ \hat{A}^{(1)}_0 \sim \hat{A}^{(4)}_0.$
From our earlier results this is equivalent to 
\beq 
 \frac{e^{-k_1 r}}{\Lambda r} \sim  \frac{1}{\Lambda^2 r^5}, 
\ \frac{1}{\Lambda^4 r^9}, 
\eeq
so a new length scale appears at ${r} \sim  \log{\Lambda},$
or more generally  ${r} \sim  L\log{\Lambda},$
if we reinstate the scale $L.$  
Note that this is a large scale for large $\Lambda.$
It is the scale beyond which the asymptotic fields, of the form
(\ref{nontail}), are applicable to describe the tail of the soliton. 

It is common to define the size of a soliton's core by reference to 
the region beyond which the fields of the asymptotic tail provide a good 
approximation to the fields of the soliton. If such a definition is used
then the soliton is large at large 't Hooft coupling.
This is in stark contrast to the commonly stated result that
the soliton has a small size, which results by defining the size 
via comparison with the approximate self-dual instanton. 
The size of the \ss soliton is therefore a more complicated issue than 
previously realized. 

In summary, there are three important scales in the problem.
The scale of the self-dual instanton, $L/\sqrt{\Lambda}$, 
the radius of curvature $L$, and the new scale of order 
$L \log{\Lambda}$. The various approximations discussed in this paper 
are valid in different regions, some of which are contiguous and therefore
allow the different approximations to be related.
These different regions correspond to the treatment of space
as flat or curved and the treatment of the partial differential equations
as linear or nonlinear.
Schematically, we may summarise the situation as:
\bea
\label{regimeslargelambda}
0< & \rho & \lesssim L/\sqrt{\Lambda},    \   \qquad  \qquad    
\mbox{\rm flat and nonlinear}       \nonumber \\
  L/\sqrt{\Lambda}   \lesssim  &\rho&  \lesssim L,   \ \qquad   \qquad   \qquad   
\mbox{\rm  flat and linear}   \nonumber \\
L \lesssim &\rho&  \lesssim  L \log{ \Lambda},         \qquad       \qquad   
\mbox{\rm   curved and linear}            \nonumber \\
  L \log{ \Lambda} \lesssim &\rho&  \phantom{\lesssim \infty} \, \  \ \ \ \ \ \   \qquad  
     \qquad   \mbox{\rm  curved and nonlinear.}
\eea
The appearance of the final region is a slightly unusual 
feature due to the fact that at large radius
nonlinear terms dominate over linear terms, despite the fact
that these terms are small.
This has led to some confusion by previous authors, who have 
incorrectly assumed the more generic behaviour that when 
functions become small the system enters a linear regime. 

\subsection{The Cherman-Ishii expansion}\quad\label{sec-CI}
Cherman and Ishii 
\cite{Cherman:2011ve}
have performed a large $r$
expansion to obtain the asymptotic fields of the \ss soliton, based
on a method first applied in a different holographic model
\cite{Panico:2008it}. 
They found that the fields have an algebraic decay with a form that
satisfies the model independent form factor relations described in
\cite{Cherman:2009gb}. However, they were only able to implement 
their approach by introducing a UV cutoff and the limit as this
cutoff is removed is problematic: prompting them to 
speculate on possible resolutions that include 
holographic renormalization and boundary counterterms.
In this subsection we describe the relation between our asymptotic
fields and those of the Cherman-Ishii expansion.
Although the Cherman-Ishii fields appear to have a more complicated
form than our expressions, we shall show that they are gauge equivalent.
Moreover, we shall see that their required UV cutoff is merely a gauge   
artifact that is a consequence of a gauge choice that is incompatible
with the holographic boundary conditions.
Although our asymptotic expansion turns out to be equivalent to the 
Cherman-Ishii expansion, our derivation has the advantage that 
the constant appearing in the expansion is directly related to the  
self-dual instanton, whereas it appears simply as an unknown constant
in the  Cherman-Ishii 
expansion, even after a gauge transformation
to remove the spurious UV cutoff.

First we highlight the relevant issue concerning the choice of gauge. 
The required condition at the conformal boundary, 
that the field strength has vanishing parallel components, is gauge invariant. 
In the AdS/CFT dictionary the gauge potential $A_{i}$ at the 
boundary corresponds to the source for the related current. 
If $F_{ij}$ is zero at the boundary, then it is always possible to choose a 
gauge in which $A_{i}$ is also set to zero at the boundary 
(so that the sources vanish). 
The chosen gauge for our expansion is already in this form. 
The starting point of our linear expansion (\ref{newlinearstart}) 
has $A_i^{(1)}=0$, moreover none of the higher order terms give a contribution
at the boundary, so $A_i$ vanishes there. 
This is why we have no need for a UV cutoff.

The expansion strategy followed in \cite{Cherman:2011ve} has led to
some confusion about the choice of gauge because they 
directly perform a radial expansion in $1/r$ and do not consider 
an expansion in $1/\Lambda$.
They  start their expansion with a dominant term at large $r$ 
given by 
\beq
\label{CI}
A_{z} =       \beta    \frac{\sigma_{j}\hat{x}_j}{r^2}+\ldots
\eeq
where $\beta$ is a constant that is left arbitrary and cannot be determined
using their approach.
The crucial point here is that this term is independent of $z$. 
All the other terms in the $1/r$ expansion are then derived on top of this 
one, choosing step by step a gauge in which $A_i=0$ at the boundary.
However, by postulating the leading term (\ref{CI}) this
implicitly contains a gauge choice, which is not necessarily
 compatible with the choice of gauge in which $A_{i}=0$ at the boundary. 
In fact it turns out that, generically, the only way to have $A_{i}$ 
vanishing at the boundary is to introduce a fictitious UV cutoff. 
The need for a cutoff simply reflects the fact that the gauge
implicitly chosen by the starting point (\ref{CI})
 is not a good one.

To relate our expansion to that of Cherman and Ishii we shall take
our leading order result, given by the pion tail (\ref{newlinearstart}),
and attempt to convert it to a gauge in which (\ref{CI}) holds.
The first step is to perform the gauge transformation given by
\be
G_1 = \exp{\left( -\frac{i}{2} \xi_0 \mu^2 \frac{\sigma_{j}\hat{x}_j}{r^2}\psi_{0}( z) \right)},
\ee 
which results in
\be
\label{piongaugetransformed}
A_z^{(1)} = 0, \qquad\qquad
A_i^{(1)} = \frac{\xi_0 \mu^2}{2r^3} 
(\sigma_{i} - 3  \sigma_j \hat{x}_i \hat{x}_j)\psi_{0}( z),
\ee
so that the pion tail is transferred entirely into the $A_i$ component.

This is a perfectly legitimate gauge, but it does not have vanishing sources 
at the boundary, because $\psi_0(\infty)\ne 0.$
A way to resolve this issue is to introduce
a UV cutoff, $z_{UV},$ and perform second gauge transformation 
given by
\be
G_2=\exp{\left( \frac{i}{2} \xi_0 \mu^2 \frac{\sigma_{j}\hat{x}_j}{r^2}
\frac{z\psi_{0}( z_{UV})}{z_{UV}} \right)},
\ee
so that (\ref{piongaugetransformed}) becomes 
\be
A_z^{(1)} =  -\frac{\xi_0 \mu^2}{2} \, \frac{\sigma_{j}\hat{x}_j}{r^2} \frac{ \psi_{0}( z_{UV}) }{z_{UV}}, \qquad
A_i^{(1)} = \frac{\xi_0 \mu^2}{2r^3} 
(\sigma_{i} - 3  \sigma_j \hat{x}_i \hat{x}_j)
\bigg(\psi_{0}( z)-\frac{z\psi_{0}( z_{UV})}{z_{UV}} \bigg).
\ee
Now $A_i^{(1)}$ vanishes at the UV boundary $z=z_{UV}$ and $A_z^{(1)}$ has the
form (\ref{CI}) with 
$\beta = -\xi_0 \mu^2 \psi_{0}( z_{UV})/(2z_{UV}).$ 
This demonstrates the equivalence between the 
Cherman-Ishii expansion and our simpler version (\ref{newlinearstart}) and
explains how the UV cutoff terms are simply gauge artefacts.

The conclusion is that the correct gauge to have vanishing sources 
is in fact (\ref{newlinearstart}) with no term like (\ref{CI}). 
In the Cherman-Ishii expansion 
there are other terms, of higher order in $1/r$ and linear in $1/\Lambda$, 
that can also be 
removed by a gauge transformation, leaving the 
physical terms that correspond to our second order and fourth order terms 
(\ref{secondorderterm}) and (\ref{azerofour2}). 
In particular, formula (\ref{azerofour2}) for 
the leading order large $r$ behaviour of $\hat A_0,$ 
does not depend on the choice of gauge for the non-abelian fields
and coincides with the one given in
\cite{Colangelo:2013pxk}, which is a correction of the
expression in \cite{Cherman:2011ve} (as this contains an error).

In the recent preprint \cite{Colangelo:2013pxk},
which appeared on the arXiv during the preparation of this manuscript,
 it is argued that the UV cutoff is a kind of coordinate singularity that 
can be removed by a very specific change of variable for the holographic 
coordinate. However, the interpretation as a coordinate singularity
is not the underlying explanation but is a pure coincidence, as follows. 
In the right gauge, the correct starting point 
for the expansion of $A_z$ is $A_z^{(1)}$ 
given by  (\ref{newlinearstart}), 
which has a $z$ dependence proportional to $\phi_0(z).$ 
The specific change of coordinate identified in \cite{Colangelo:2013pxk}
is to use $\tilde z=\psi_0(z)$ as the holographic variable.
The correct dependence of $A_z^{(1)}$ 
maps to an $A_{\tilde z}^{(1)}$ that is independent
of $\tilde z,$ and hence is of the form (\ref{CI}), simply because this
specific choice of $\tilde z$ obeys $d\tilde z/dz=\phi_0,$
which cancels the $z$-dependent factor of $\phi_0$ in $A_z.$ 
Independence of the holographic coordinate happens only for this 
specific choice of coordinate and for a generic coordinate 
the correct formula is obtained from (\ref{newlinearstart}).

\section{Numerical computations}\quad
\label{numerics}
In this section we describe our numerical scheme for the computation
of the \ss soliton, together with some of the results it generates.
We shall see that the numerical results are in good agreement with
the analytical approximations discussed in the previous sections.
 
Static $SO(3)$ symmetric fields have the form
\cite{Witten:1976ck,Forgacs:1979zs}

\be
A_j=\bigg(\frac{1+\Phi_2}{r}\varepsilon_{jak}\hat x_k
+\frac{\Phi_1}{r}(\delta_{ja}-\hat x_j\hat x_a)+a_r{\hat x_j\hat x_a}
\bigg)\frac{\sigma_a}{2},
\quad
A_z=a_z\hat x_a\frac{\sigma_a}{2}, \quad \widehat A_0,
\ee
where the fields $\Phi_1,\Phi_2,a_r,a_z,\hat A_0$ are functions of $r$ and $z.$

Writing $\Phi=\Phi_1+i\Phi_2,$ \
$f_{rz}=\partial_r a_z-\partial_z a_r$
 and $D_r\Phi=\partial_r\Phi-ia_r\Phi,$  
the expression for the baryon number becomes
\be
B=-
\int_0^\infty dr \int_{-\infty}^\infty dz\ 
\frac{1}{2\pi}\bigg\{f_{rz}(1-|\Phi|^2)+
i(D_r\Phi \overline{D_z\Phi}-\overline{D_r\Phi} D_z\Phi)\bigg\}.
\ee
In terms of these variables, the energy obtained from the 
action (\ref{staticaction}) has three terms,
 $E=4\pi(E_{SU(2)}+E_{U(1)}+E_{CS}),$ where
\be
E_{SU(2)}=
\int_0^\infty dr \int_{-\infty}^\infty dz\ 
\bigg\{
\h|D_r\Phi|^2+\k|D_z\Phi|^2+\frac{r^2\k}{2}f_{rz}^2+\frac{\h}{2r^2}(1-|\Phi|^2)^2
\bigg\},
\label{esu2}
\ee
\be
E_{U(1)}=-
\int_0^\infty dr \int_{-\infty}^\infty dz\ 
\bigg\{
\frac{1}{2}r^2\bigg(\h(\partial_r \hat A_0)^2+\k(\partial_z \hat A_0)^2\bigg)
\bigg\},
\label{eu1}
\ee
\be
E_{CS}=
-\frac{1}{\Lambda}\int_0^\infty dr \int_{-\infty}^\infty dz\ 
\bigg\{
4\hat A_0
\bigg(f_{rz}(1-|\Phi|^2)+
i(D_r\Phi \overline{D_z\Phi}-\overline{D_r\Phi} D_z\Phi)\bigg)
\bigg\}.
\label{ecs}
\ee
For reference, the flat space self-dual instanton is given by
\be
\Phi=\frac{2rz+i(r^2-z^2-\mu^2)}{\rho^2+\mu^2}, \quad
a_r=\frac{2z}{\rho^2+\mu^2}, \quad
a_z=\frac{-2r}{\rho^2+\mu^2},
\label{selfdualvortex}
\ee
where, as earlier, $\rho^2=r^2+z^2.$
The required soliton has $B=1$ and is a vortex in the reduced theory on the 
half-plane $r\ge 0.$  On the boundary $\{r=0\} \cup \{\rho=\infty\}$
the complex field $\Phi$ has unit modulus
and its phase varies by $2\pi$ around the boundary. 
Setting $\mu=0$ in (\ref{selfdualvortex}) gives the fields
\be
\Phi=\frac{2rz+i(r^2-z^2)}{\rho^2}, \qquad
a_r=\frac{2z}{\rho^2}, \qquad
a_z=\frac{-2r}{\rho^2}, \qquad
\label{singular}
\ee
which are pure gauge but have a singularity at 
the point $\rho=0.$
These fields satisfy $|\Phi|=1$ and
$D_r\Phi=D_z\Phi=f_{rz}=0,$ which
are the natural boundary conditions to impose
as $\rho\rightarrow \infty.$
In particular, the phase of $\Phi$ varies by $2\pi$ along
this boundary. 
The boundary conditions along the line $r=0$ are
$\Phi=-i, \ D_r\Phi=D_z\Phi=0,$
which are those of the finite size self-dual instanton. 
A series expansion of the field equations around $r=0$ confirms
that these are the correct boundary conditions as $r\rightarrow 0,$ 
together with $\partial_r \hat A_0=0.$ 
In summary, the boundary conditions at $r=0$ are given by
\be
\Phi=-i, \quad a_r=\partial_r\Phi_1, \quad a_z=0, 
\quad \partial_r \hat A_0=0,
\ee
and as $\rho\rightarrow \infty$ the fields are given by (\ref{singular})
together with $\hat A_0\rightarrow 0.$

The field equations that follow from the variation of the energy $E$
are solved using a heat flow method. For the fields $\Phi_1,\Phi_2,a_r,a_z$
this corresponds to gradient flow associated with the energy
$E_{SU(2)}+E_{CS},$ in the Coulomb gauge $\partial_r a_r+\partial_z a_z=0.$
For $\hat A_0$ the heat flow corresponds to gradient flow
associated with the energy $-E_{U(1)}-E_{CS},$  where the negative signs
are due to the negative sign that appears in front of the energy
(\ref{eu1}), arising because $\hat A_0$ is the time component of a 
gauge potential.
The problem may be viewed as a constrained energy minimization, where
the energy $E_{SU(2)}+E_{CS}$ is to be minimized subject to the constraint
that $\hat A_0$ satisfies the field equation
\be
\frac{1}{r^2H^{1/2}}\partial_r(r^2\partial_r \hat A_0)+\partial_z(H^{3/2}\partial_z \hat A_0)
=\frac{4}{\Lambda r^2}  
\bigg\{f_{rz}(1-|\Phi|^2)+
i(D_r\Phi \overline{D_z\Phi}-\overline{D_r\Phi} D_z\Phi)\bigg\},
\label{seqn}
\ee
which is a curved space Poisson equation sourced by the instanton charge
 density.

As an initial condition for the numerical relaxation the self-dual 
instanton fields (\ref{selfdualvortex}) are taken with a spatially dependent 
size $\mu(r,z)$ so that $\mu(0,0)\ne 0$ 
but $\mu(r,z)=0$ for sufficiently large $\rho.$  
For $\hat A_0$ the initial condition is that it vanishes everywhere.

As we have seen from the analysis in the previous sections,
and will be confirmed by the numerical computations in this section,
the fields decay more slowly in the $z$ direction
than in the $r$ direction, due to the warped metric. 
As a result, it turns out to be computationally efficient to perform the 
change of variable $z=\tan w,$ so that the infinite domain of $z$ 
transforms to the finite interval $w\in[-\frac{\pi}{2},\frac{\pi}{2}].$
At the boundaries $w=\pm \frac{\pi}{2}$ the fields (\ref{singular})
now give the boundary conditions 
$\ \Phi=-i, \ a_r=a_z=\hat A_0=0.$

The numerical solution is computed on a grid with a boundary at
a finite value $r=r_\star.$ 
The boundary conditions applied at this
simulation boundary are that the fields are given by 
the pure gauge fields (\ref{singular}) 
together with $\hat A_0=0,$ that is,
\be
\Phi=\frac{2r_\star \tan w+i(r_\star^2-\tan^2w)}
{r_\star^2+\tan^2w}, \quad
a_r=\frac{2\tan w}{r_\star^2+\tan^2w}, \quad
a_z=-\frac{2r_\star}{r_\star^2+\tan^2w}, \
\hat A_0=0.
\label{bcrmax}
\ee
Note that the $2\pi$ phase winding of $\Phi$ now takes place
along the single boundary $r=r_\star.$
It has been verified that the solutions are
insensitive to the choice of this finite boundary, 
providing $r_\star$ is taken to be sufficiently large.
The simulation details depend upon the value of $\Lambda,$ as this
sets the scale of the soliton, but for $\Lambda$ of order one
a typical grid contains $400\times 200$ points in the $(r,w)$-plane
with $r_\star=40.$

To display the results of the numerical computations it is convenient
to plot the abelian potential $\hat A_0,$ as this is a scalar 
quantity that is invariant under $SU(2)$ gauge transformations, and
in addition 
the $U(1)$ gauge freedom is fixed by our earlier prescription that
$\hat A_I=0$ and $\hat A_0\to 0$ as $\rho\to\infty.$
Furthermore, we have the simple explicit 
expression (\ref{sda0}) for $\hat A_0$ 
within the flat space self-dual approximation,
that can be used to compare to the numerical result. 
\begin{figure}
\includegraphics[width=10cm]{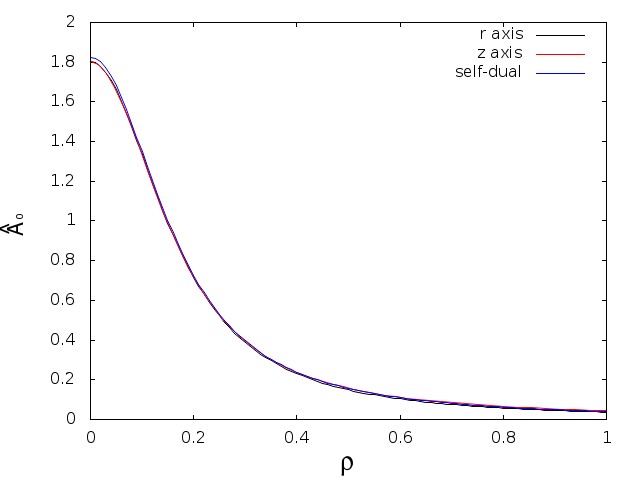}
\includegraphics[width=6cm]{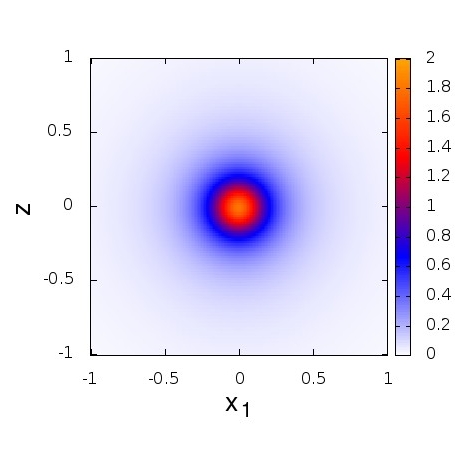}  
\caption{
The abelian potential $\widehat A_0$ 
for the soliton with $\Lambda=200.$
The left image displays plots of  
$\widehat A_0$ 
along the $r$-axis (black curve)
and the $z$-axis (red curve). The flat space self-dual approximation
(blue curve) is included for comparison.
All three curves are almost indistinguishable
as the self-dual field provides a good approximation in this range, 
apart from a very slight overshoot at the origin.
The right image is a plot of  $\widehat A_0$ in the plane $x_2=x_3=0$
and demonstrates the approximate $SO(4)$ symmetry in this region.
}\label{fig-core001} \end{figure}
\begin{figure}
\begin{center}
\includegraphics[width=10cm]{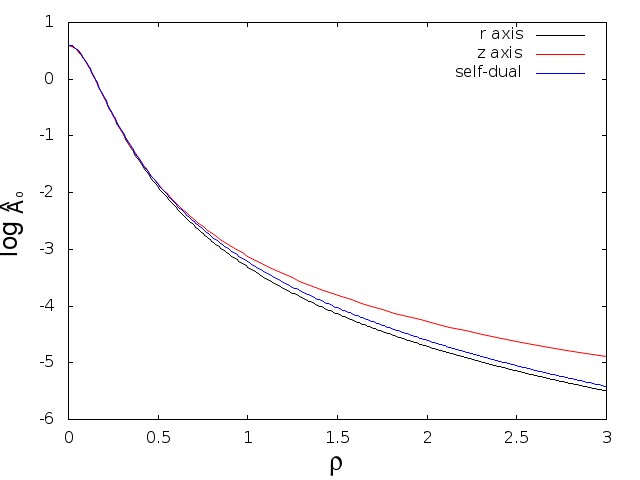}
\caption{
For the soliton with $\Lambda=200,$ 
the plot displays $\log \widehat A_0$ against $\rho$ 
along the $r$-axis (black curve)
and the $z$-axis (red curve). The flat space self-dual approximation
(blue curve) is included for comparison.
There is a faster decay along the $r$-axis than along the $z$-axis.
}\label{fig-tail001} 
\end{center}
\end{figure}

We first compute the soliton for a large value of $\Lambda,$ where we
expect the self-dual instanton to be a good approximation, at least in
the region $\rho\lesssim 1.$ 
Figure~\ref{fig-core001} displays a plot of  
$\hat A_0$ for the value $\Lambda=200.$ 
The plot in the left image presents $\hat A_0$ along the $r$ and $z$ axes,
together with the $SO(4)$ symmetric self-dual instanton approximation 
(\ref{sda0}) with the instanton size given by (\ref{size}).
All three curves are almost indistinguishable, which confirms that the
the self-dual instanton provides a good approximation in this range, 
for this large value of $\Lambda.$
The plot in the right image presents $\hat A_0$ in the plane $x_2=x_3=0,$
and demonstrates the approximate $SO(4)$ symmetry for $\rho\lesssim 1.$
To see a deviation from the self-dual approximation requires 
an examination of the region $\rho>1.$ As $\hat A_0$ is small in this region
then the appropriate quantity to plot is $\log\hat A_0,$ which is
presented in Figure~\ref{fig-tail001} for $0\le\rho\le 3.$
The lack of $SO(4)$ symmetry is now more apparent, with a slower decay
along the $z$-axis than along the $r$-axis, as predicted by the 
analytic calculations.
\begin{figure}
\includegraphics[width=10cm]{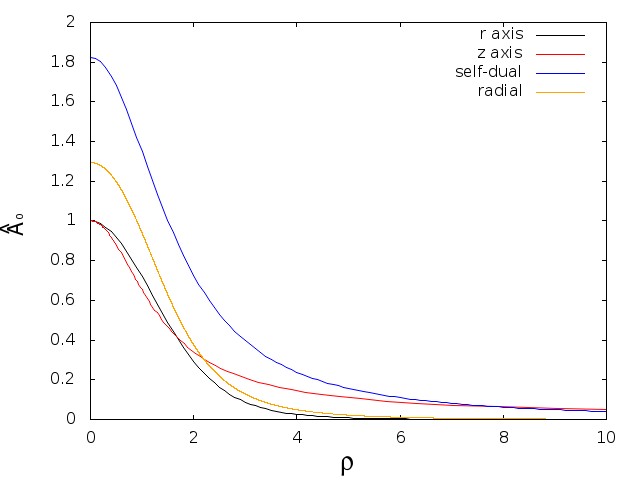}
\includegraphics[width=6cm]{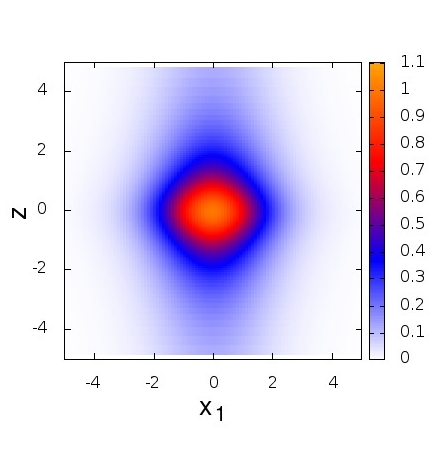}  
\caption{
The abelian potential $\widehat A_0$ 
for the soliton with $\Lambda=2.$
The left image displays plots of  
$\widehat A_0$ 
along the $r$-axis (black curve)
and the $z$-axis (red curve). The flat space self-dual approximation
(blue curve) is included for comparison, 
together with the radial approximation (orange curve).
Note the faster decay along the $r$-axis than along the $z$-axis.
The right image is a plot of  $\widehat A_0$ in the plane $x_2=x_3=0.$
}\label{fig-core1} \end{figure}
\begin{figure}
\begin{center}
\includegraphics[width=10cm]{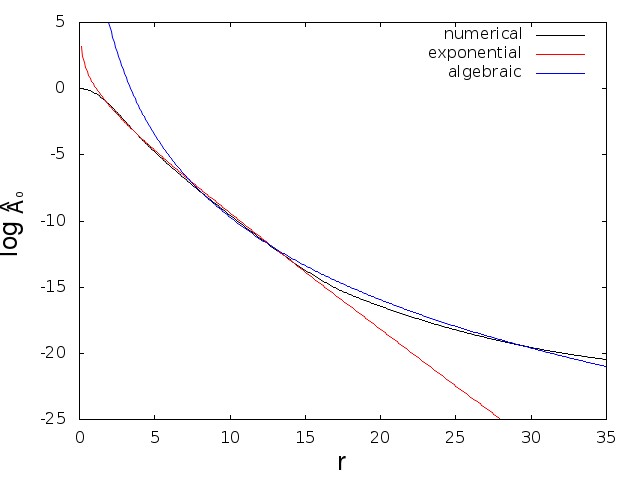}
\caption{
For the soliton with $\Lambda=2,$ 
the plot displays $\log \widehat A_0$ against $r$ 
along the $r$-axis (black curve). 
The red curve is the exponential decay predicted by the linear
approximation in curved space and the blue curve is the algebraic
decay predicted by the nonlinear approximation in curved space.
Exponential decay is a good approximation in the region
$1\lesssim r \lesssim 15$ 
and algebraic decay is a good approximation in the region
$r \gtrsim 8$ 
}\label{fig-tail1} 
\end{center}
\end{figure}

To see a demonstrable difference between the self-dual 
instanton and the numerical solution requires a value
of $\Lambda$ that is of order one. This is also the case if we are to
provide numerical evidence to support our analytic calculations
concerning the applicability of the linear and nonlinear descriptions
of the soliton tail in different regions. The most relevant regime
from the physical point of view is large $\Lambda,$ but as we have seen,
the three length scales involved are of order $1/\sqrt{\Lambda},1,\log\Lambda.$
For large $\Lambda$ this gives a separation of scales that is 
difficult to encompass within a single simulation.  
By going to parameter values of $\Lambda$ that are of order one, we can
bring these three length scales closer together, so that all three are
simultaneously accessible within a feasible simulation.
 
Figure~\ref{fig-core1} displays a plot of  
$\hat A_0$ for the value $\Lambda=2.$ 
The plot in the left image presents $\hat A_0$ along the $r$ and $z$ axes,
together with the self-dual instanton approximation and the radial
approximation described in section \ref{radialandselfdual}.
The slower decay along the $z$ axis than along the $r$ axis is now
clearly visible. The self-dual instanton is a poor approximation for
this value of $\Lambda,$ even for small $\rho.$ The radial approximation
improves on the self-dual approximation, but there is still a considerable
error, as expected from an approximation that assumes $SO(4)$ symmetry.
The plot in the right image presents $\hat A_0$ in the plane $x_2=x_3=0,$
and clearly displays the lack of $SO(4)$ symmetry. The abelian 
potential is stretched out along the $z$ direction, 
corresponding to the slower rate of decay along the holographic direction,
in agreement with the earlier analysis.

To examine the soliton tail, we plot $\log\hat A_0$ 
against $r$ (along the $r$-axis) in Figure~\ref{fig-tail1}. 
Also included in this plot is the leading order exponential decay 
predicted by the linear analysis, namely 
$\hat A_0={\alpha_1e^{-k_1r}}/{r},$
and the leading order algebraic decay predicted by the nonlinear analysis,
$\hat A_0={\alpha_2}/{r^9},$ where $\alpha_{1,2}$ are constants. 
It can be seen that exponential decay is a good approximation in the region
$1\lesssim r \lesssim 15,$  where the linear regime is valid, and
algebraic decay is a good approximation in the region $r \gtrsim 8,$
which is the nonlinear regime. The slight discrepancy between the algebraic
form and the numerical result at large $r$ is due to the finite boundary
at $r=r_\star=40,$ which is not far beyond the range plotted in this figure.

In summary, the numerical results presented in this section demonstrate
that the flat space self-dual instanton is a good approximation to
the \ss soliton for $\rho\lesssim L,$ providing $\Lambda$ is large.
Furthermore, we have provided numerical evidence to
support the analytic results obtained in this paper regarding the validity
of the linear approximation immediately outside the soliton core,
together with its breakdown at large scales, 
where nonlinear terms are dominant. 

\section{Conclusion}\quad
\label{conclusion}
Using a combination of analytic and numerical methods we have investigated
the properties of the \ss soliton, together with a range of approximations
that have been applied to study this soliton. We have determined the regimes
of validity of these approximations and shown how they may be related in
regions where they overlap. This analysis has 
clarified the source of some contradictory results 
in the literature and resolved some outstanding
issues, including the applicability of the flat space
self-dual instanton, the detailed properties of the asymptotic soliton tail,
and the role of the UV cutoff required in previous investigations.
We have shown how to relate the asymptotic fields to the self-dual 
instanton description valid at the core, and revealed the existence
of a new large scale, that grows logarithmically with the 't Hooft
coupling, at which the soliton fields enter a nonlinear regime.

The leading order term in the soliton tail is provided by the
massless pion field and the classical inter-soliton force has the
same structure as in the Skyrme model. Hence there should be an
attractive channel that leads to classical multi-soliton bound states
that can be quantized within a collective coordinate approximation
to provide holographic nuclei. However, these multi-solitons are not 
expected to have the $SO(3)$ symmetry of the single soliton, so 
it will be a significant computational challenge to construct these
solutions numerically.

\section*{Acknowledgements}
This work is funded by the EPSRC grant EP/K003453/1 and 
the STFC 
grant ST/J000426/1.
We thank Aleksey Cherman, Kasper Peeters and Marija Zamaklar
for useful discussions.


\begin{thebibliography}{11}

\bibitem{Atiyah:1989dq}
  M.~F.~Atiyah and N.~S.~Manton,
  ``Skyrmions From Instantons,''
  Phys.\ Lett.\ B {\bf 222} (1989) 438.


\bibitem{Sutcliffe:2010et}
  P.~Sutcliffe,
  JHEP {\bf 1008} (2010) 019
  [arXiv:1003.0023 [hep-th]].



\bibitem{Sakai:2004cn}
  T.~Sakai and S.~Sugimoto,
  ``Low energy hadron physics in holographic QCD,''
  Prog.\ Theor.\ Phys.\  {\bf 113} (2005) 843
  [hep-th/0412141].

\bibitem{Sakai:2005yt}
  T.~Sakai and S.~Sugimoto,
  ``More on a holographic dual of QCD,''
  Prog.\ Theor.\ Phys.\  {\bf 114} (2005) 1083
  [hep-th/0507073].

\bibitem{Hong:2007kx}
  D.~K.~Hong, M.~Rho, H.~-U.~Yee and P.~Yi,
  ``Chiral Dynamics of Baryons from String Theory,''
  Phys.\ Rev.\ D {\bf 76} (2007) 061901
  [hep-th/0701276 [HEP-TH]].

\bibitem{Hata:2007mb}
  H.~Hata, T.~Sakai, S.~Sugimoto and S.~Yamato,
  ``Baryons from instantons in holographic QCD,''
  Prog.\ Theor.\ Phys.\  {\bf 117} (2007) 1157
  [hep-th/0701280 [HEP-TH]].

\bibitem{Hashimoto:2008zw}
  K.~Hashimoto, T.~Sakai and S.~Sugimoto,
  ``Holographic Baryons: Static Properties and Form Factors from Gauge/String Duality,''
  Prog.\ Theor.\ Phys.\  {\bf 120} (2008) 1093
  [arXiv:0806.3122 [hep-th]].


\bibitem{Hong:2007dq}
  D.~K.~Hong, M.~Rho, H.~-U.~Yee and P.~Yi,
  ``Nucleon form-factors and hidden symmetry in holographic QCD,''
  Phys.\ Rev.\ D {\bf 77} (2008) 014030
  [arXiv:0710.4615 [hep-ph]].


\bibitem{Kim:2008pw}
  K.~-Y.~Kim and I.~Zahed,
  JHEP {\bf 0809} (2008) 007
  [arXiv:0807.0033 [hep-th]].
  

\bibitem{Pomarol:2008aa}
  A.~Pomarol and A.~Wulzer,
  ``Baryon Physics in Holographic QCD,''
  Nucl.\ Phys.\ B {\bf 809} (2009) 347
  [arXiv:0807.0316 [hep-ph]].

\bibitem{Panico:2008it}
  G.~Panico and A.~Wulzer,
  ``Nucleon Form Factors from 5D Skyrmions,''
  Nucl.\ Phys.\ A {\bf 825} (2009) 91
  [arXiv:0811.2211 [hep-ph]].

\bibitem{Cherman:2009gb}
  A.~Cherman, T.~D.~Cohen and M.~Nielsen,
  ``Model Independent Tests of Skyrmions and Their Holographic Cousins,''
  Phys.\ Rev.\ Lett.\  {\bf 103} (2009) 022001
  [arXiv:0903.2662 [hep-ph]].

\bibitem{Cherman:2011ve}
  A.~Cherman and T.~Ishii,
  ``Long-distance properties of baryons in the Sakai-Sugimoto model,''
  Phys.\ Rev.\ D {\bf 86} (2012) 045011
  [arXiv:1109.4665 [hep-th]].


\bibitem{Colangelo:2013pxk}
  P.~Colangelo, J.~J.~Sanz-Cillero and F.~Zuo,
  ``Large-distance properties of holographic baryons,''
  arXiv:1306.6460 [hep-ph].



\bibitem{Witten:1976ck}
  E.~Witten,
  ``Some Exact Multi - Instanton Solutions of Classical Yang-Mills Theory,''
  Phys.\ Rev.\ Lett.\  {\bf 38} (1977) 121.


\bibitem{Forgacs:1979zs}
  P.~Forgacs and N.~S.~Manton,
  ``Space-Time Symmetries in Gauge Theories,''
  Commun.\ Math.\ Phys.\  {\bf 72} (1980) 15.

\end{thebibliography}
\end{document}